\documentclass[]{article} 
\usepackage[utf8]{inputenc} 

\usepackage{geometry} 
\usepackage{amsmath}
\usepackage{graphicx} 
\usepackage{color}

\usepackage{booktabs} 
\usepackage{array} 
\usepackage{paralist} 
\usepackage{verbatim} 
\usepackage{subfig} 
\usepackage{tikz}
\usepackage{tkz-euclide}

\usepackage{fancyhdr} 
\usepackage{sectsty}
\allsectionsfont{\sffamily\mdseries\upshape} 

\usepackage{amsfonts}

\usepackage{amsmath}

\begin{document}




\title{Drawing maps on oriented surfaces}
\author{Gunnar Brinkmann}

\maketitle

\begin{abstract}
  In this article we describe a program -- called {\em planar\_draw} -- to draw maps on oriented surfaces in the plane. The drawings are coded as tikz
  files that can easily be manipulated and used in latex documents.  Next to plane maps -- a case for which already several programs exist -- the program allows
  to draw maps of genus at least one inside a fundamental polygon or with non-contractible cycles displayed as disjoint cycles that have to be identified.
  Several options allow to tailor the output for individual needs -- e.g.\ by forcing some edges to be completely inside the fundamental polygon.  In
  combination with a program embedding graphs, the tool can also be used for graphs that do not already come with an embedding in an orientable surface.
  
\end{abstract}


\section{Introduction}

{\em Drawing} non-planar maps means to find a flat representation, which can be printed on a sheet of paper. This implies either crossing edges or edges divided into parts where the endpoints of these parts
have to be identified. In mathematics, a well accepted way to represent oriented 2-manifolds is by a fundamental polygon. A drawing in a fundamental polygon can e.g. be obtained by cutting a surface
of genus $g$ open along $2g$ (well chosen) cycles sharing one common point. This results in an outer face with $4g$ sides with the whole map in the interior or on the boundary. In order to obtain the original map,
sides coming from the same cycle must be identified in a prescribed way. Edges crossing the cut cycles are drawn in parts.

Another way to draw non-planar maps of genus $g$ is to cut them open along $g$ (well chosen) disjoint cycles. This results in plane maps with $2g$ special faces which describe the cut cycles and have to be connected by tubes or identified in order
to obtain the original surface and map.

In both cases we consider two properties to be essential:

\begin{description}
\item[(i)] The vertices should be sufficiently separated from each other. This is not an easy requirement in case of many vertices.
\item[(ii)] It should be easy to see the rotational order of neighbours of each vertex. In this sense the drawing is in a sense monotonically better than a list of neighbours, which is often used to describe maps.
\end{description}

Furthermore there is one more requirement that is maybe not essential but at least beneficial:

\begin{description}
\item[(iii)] One should be able to easily recognize as many faces of the map as possible.
\end{description}

There are articles with theoretical results about desirable properties of such drawings and also algorithms for drawing maps,
e.g.\ \cite{drawing_genus_2009},\cite{arjana_draw}. Nevertheless these articles do not come with software that actually allows to draw the maps. In \cite{torusdraw} results about drawing toroidal maps and an algorithm to
produce such drawings are given and the algorithm is implemented in the software package {\em Groups and Graphs}.
The intention of this article is not to present new techniques or new theoretical insights into drawing maps on oriented surfaces, but to present a {\bf tool} for drawings maps
that will hopefully be useful for researchers, save them a lot of time preparing their papers, and help them in research. 
Planar\_draw can be downloaded from
\verb+http://twicaagt.ugent.be/~gbrinkma/planar_draw.html+.

For a set $E$ of undirected edges $\{x,y\}$, the set $E_o$ of oriented edges is the set containing for each edge $\{x,y\}\in E$ the {\em oriented edges} $[x,y]$ starting at $x$ and $[x,y]^{-1}=[y,x]$ starting at $y$.
A {\em combinatorial map} on an orientable surface is a graph $G=(V,E)$ together with a function $n(): E_o\to E_o$ defining for each $v\in V$ a rotational order of oriented edges $[v,.]$ considered as clockwise.
Defining the relation $[v,x]\equiv n([x,v])$ -- in this case $[v,x],n([x,v])$ is sometimes called an {\em angle} -- the equivalence classes generated by $\equiv$ are called {\em faces} and with $F$ the set
of all faces, the number $g$ satisfying $|V|-|E|+|F|=2-2g$ is called the genus of the combinatorial map. The equivalence between these combinatorial concepts and the toplogical concept of graphs embedded on orientable compact 2-manifolds
is e.g.\ explained in \cite{top_graph_theory}. When referring to graphs that do not come with an embedding in a surface, we use the term {\em abstract graph} to emphasize this property.

A combinatorial map of genus $g$ can be drawn inside a fundamental polygon for genus $g$ without crossing edges (but with edges crossing the boundary of the
fundamental polygon).  Depending on requirements on the fundamental polygon or the drawing, really producing such a drawing can be extremely hard and time consuming by hand.
As also mentioned in \cite{arjana_draw}, it is e.g.\ easy to (theoretically) produce drawings in fundamental polygons where all edges of the fundamental
polygon are edges of the graph and also drawings where all vertices are inside the fundamental polygon and edges cross the boundary of the fundamental polygon
at most once. The first property implies that vertices occur more than once on the boundary, which is difficult to combine with requirement (ii).
At first sight, the second option sounds very nice, but it also comes at a high price: the standard fundamental polygon of a surface of genus $g$ has $4g$
sides, but such drawings might need fundamental polygons with many more sides.

The requirements for planar\_draw are:

\begin{description}
\item[(a)]  The program must be publicly available, free, and a stand alone program outside of large software packages.
\item[(b)]  It must use an easy code for the input maps, that can be generated by other programs or by hand.
\item[(c)]  It must be efficient enough to handle {\em reasonably sized} maps of not too large genus.
\item[(d)]  It must produce output that can still be modified and adapted to individual needs -- e.g. by moving vertices, adding labels, changing sizes, etc.
\end{description}

Requirement (a) is realized by implementing it as a freely available standard C-program that reads from {\em stdin} and writes on {\em stdout}.

For (b), the code used is {\em
  planarcode}, a simple binary code used by several programs as e.g. {\em plantri}\cite{plantri}, {\em surftri}\cite{sulanke} and -- maybe more
important in this context as we will see -- the programs {\em multi\_genus} \cite{genuscomp} and {\em multi\_allembed} \cite{allembed} which construct maps from
abstract graphs, so that the abstract graphs can be drawn as maps. These programs are also used to construct the embeddings and numbers of embeddings of abstract graphs mentioned later
in this article.
Next to the binary code also a human readable and writable version of planarcode can be read. The vertices are always
$1,2,\dots ,|V|$. The first number of the code gives the number of
 vertices and then the lists of neighbours in rotational order follow -- each ended with a $0$. E.g.\ \verb+6 2 4 3 0+ \verb+1 5 6 0+ \verb+1 5 6 0+ \verb+1 6 5 0+ \verb+2 3 4 0+ \verb+4 2 3 0+ codes a map of $K_{3,3}$
 on the torus. So in this case, vertex $1$ is adjacent to $2,4,3$ in this order.

 For (c) it must of course be said that in general maps with, say, 100 vertices or 500 edges or with genus 12 would not allow drawings in which too much can
 be seen anyway. All running times given in this article are for a Core~i7-9700 processor restricted to 3 GhZ.
 As an example for the efficiency, planar\_draw takes about 90 seconds to draw genus 2 maps of the 250\ 812 known
 connected torus obstructions (three more known obstructions are disconnected) with between $8$ and $24$ vertices (see \cite{torusobstructions}) inside a fundamental
 polygon.  Producing such drawings of all 741 minimum genus maps (that is: maps with genus 7) of the 18 (3,9)-cages with
 58 vertices takes about $3,2$ seconds. The times are for producing the tikz code without compiling it e.g. with {\em pdflatex}.  Of course, these are just
 examples. Cases where the program takes much longer -- especially if no maps are given, but graphs that still have to be embedded -- will be described later.

 Choosing the option to draw maps not inside a fundamental polygon, but with disjoint cutting cycles (see Section \ref{sec:tubes}) takes considerably more
 time: the same 250\ 812 genus 2 maps of torus obstructions take about $34$ minutes (nevertheless still less than $0.01$ seconds per map) and the 741
 genus 7 maps of the 18 (3,9)-cages take about $45$ minutes -- in average a bit less than $4$ seconds per map. The reason for the increase in computing time will be
 discussed in Section~\ref{sec:tubes}.

 Aim (d) is realized by writing not jpeg, ps, or pdf as output, but by writing tikz code, that can easily be included in latex documents and can easily be modified by hand, e.g. by colouring edges, making undirected edges directed,
 or by changing sizes, labels, or positions of vertices.

\section{Drawing maps inside a fundamental polygon}\label{sec:fundpol}

The opinions about which properties of a drawing are important surely differ a lot, but for planar\_draw
we favor the following properties for drawings inside fundamental polygons:

\begin{itemize}
\item The drawing should be inside a fundamental polygon with the minimum number of sides, that is: $4g$ sides for a map of genus $g$.
\item All vertices should be strictly inside the fundamental polygon.
\item The unique vertex formed by the start- and end-points of all arrows on the boundary lies inside a face.
\end{itemize}

Nevertheless planar\_draw also allows the user to deviate from the second and third requirement if wished, as will be described in section~\ref{sec:alternative}.

\begin{figure}[tb]
	\centering
       \includegraphics[width=0.9\textwidth]{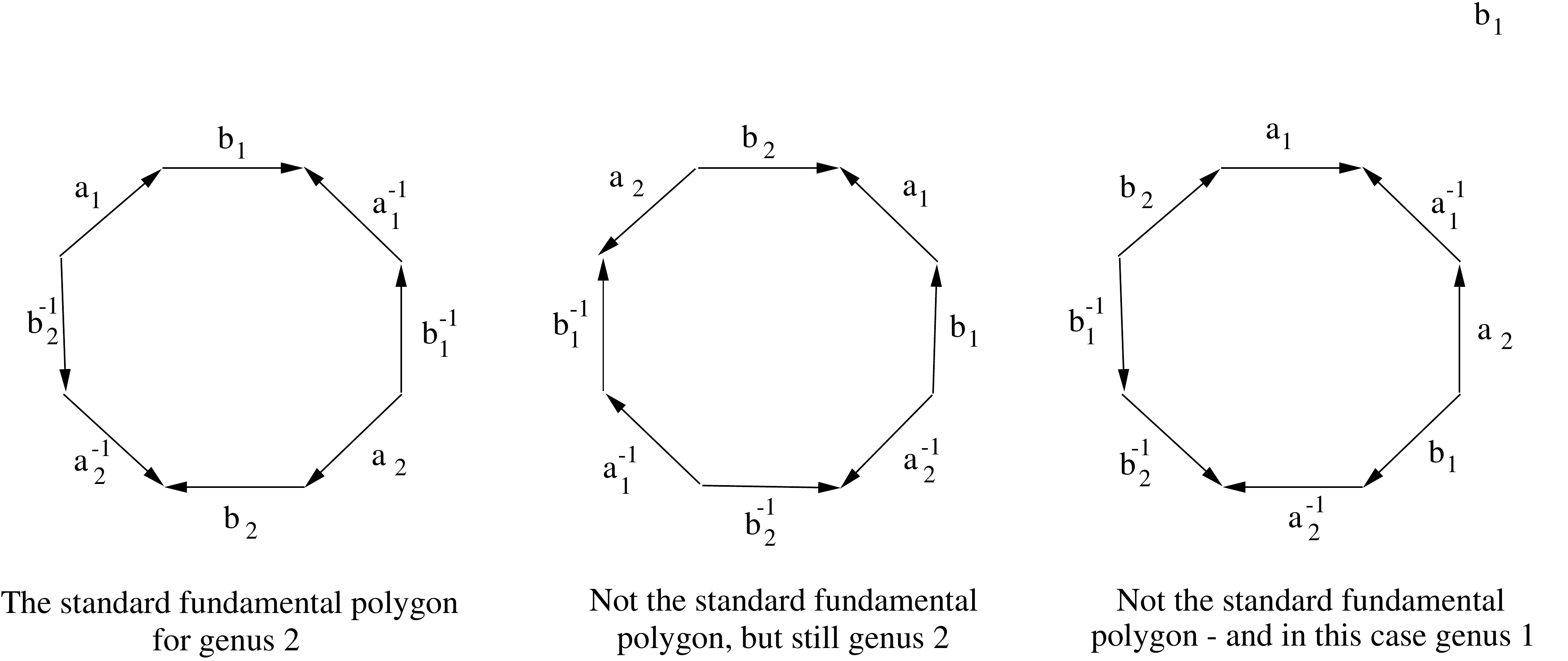}
       \caption{Three fundamental polygons with $8$ boundary edges. The exponents at the labels are in fact not necessary as the directions are also given by the arrows.}
	\label{fig:diffgen}
\end{figure}

Though we choose for a fundamental polygon with the minimum number of sides, we do not restrict ourselves to the {\em standard fundamental polygon}.  In the
standard fundamental polygon of an oriented surface of genus $g>0$, for a chosen direction of the boundary, the edges of the polygon are normally represented as
arrows and can be labeled, so that the labels are $a_1,b_1,a_1^{-1},b_1^{-1},\dots ,a_g,b_g,a_g^{-1},b_g^{-1}$ where exponent $-1$ means that the direction of
the arrow is against the chosen direction of the boundary and a missing exponent means that it is in the chosen direction of the boundary. Arrows with the same
base label have to be identified by gluing tail to tail and point to point.  In fundamental polygons with a minimum number of sides for the genus, after
identification, these arrows form non-contractible cycles of the surface. As such cycles will repeatedly be used in this article, we abbreviate non-contractible
cycles as {\em nc-cycles}.  Different orderings of arrows and labels and therefore different rules of identification of the sides of the polygon than the
standard one, can lead to surfaces of the same genus or of a smaller genus -- an example is given in Figure~\ref{fig:diffgen}.

Also our choices have some price: the start- and end-points of the boundary arrows of the fundamental polygon are one point of the surface. The default is that
it is inside a face. In a surface of genus $g$, the boundary of the face is cut by $2g$ cycles -- each cycle cutting it at least once and even at least twice in
case the boundary of the face is a simple cycle. So in a triangulation of genus $g$ the three edges of the triangle with the common vertex of the
non-contractible cycles are cut at least $4g$ times as all faces are simple cycles.  This implies that under these circumstances there is an edge being cut at least
$\lceil \frac{4g}{3}\rceil$ times. So already for triangulations of the torus at least one edge must cross the boundary of the fundamental polygon at least
twice -- and for triangulations of higher genus much more often.

\begin{figure}[tb]
	\centering
        \resizebox{0.4\textwidth}{0.4\textwidth}
        {
          \begin{tikzpicture}[scale=0.065]
\def\vertexscale{1.20}
\def\labelscale{1.40}
\node [circle,black,draw,scale=\vertexscale] (1) at (31.51581,-0.00000) {1};
\node [circle,black,draw,scale=\vertexscale] (2) at (-31.51581,0.00000) {2};
\node [circle,black,draw,scale=\vertexscale] (3) at (-0.00000,31.51581) {3};
\node [circle,black,draw,scale=\vertexscale] (4) at (0.00000,-31.51581) {4};
\tkzDefPoint(-70.71068,0.00000){5}
\tkzDefPoint(-70.71068,-70.71068){6}
\tkzDefPoint(0.00000,-70.71068){7}
\tkzDefPoint(70.71068,0.00000){8}
\tkzDefPoint(0.00000,70.71068){9}
\tkzDefPoint(-70.71068,70.71068){10}
\tkzDefPoint(70.71068,70.71068){11}
\tkzDefPoint(70.71068,-70.71068){12}

\tkzDefPoint[black,circle,draw,fill=white,scale=0.75,line width=1mm](0,0){13}
\tkzDefPoint[black,circle,draw,fill=white,scale=0.35,line width=1mm](-15.7,15.7){14}
\tkzDefPoint[black,circle,draw,fill=white,scale=0.35,line width=1mm](-15.7,-15.7){15}
\draw [green,line width=1mm] (13) to (14);
\draw [green,line width=1mm] (13) to (15);
\tkzDefPoint[](-35.4,70.7){16}
\tkzDefPoint[](-35.4,-70.7){17}
\draw [green,line width=1mm] (14) to (16);
\draw [green,line width=1mm] (15) to (17);

\tkzDefPoint[black,circle,draw,fill=white,scale=0.35,line width=1mm](15.7,15.7){19}
\tkzDefPoint[black,circle,draw,fill=white,scale=0.35,line width=1mm](15.7,-15.7){20}
\draw [purple,line width=1mm] (13) to (19);
\draw [purple,line width=1mm] (13) to (20);
\tkzDefPoint[](35.4,70.7){21}
\tkzDefPoint[](35.4,-70.7){22}
\draw [purple,line width=1mm] (19) to (21);
\draw [purple,line width=1mm] (20) to (22);
\node [black,circle,draw,fill=white,scale=0.75,line width=1mm] (13) at (0.0,0.0) {};

\draw [black] (1) to (8);
\draw [black] (1) to (4);
\draw [black] (1) to (3);
\draw [black] (2) to (5);
\draw [black] (2) to (3);
\draw [black] (2) to (4);
\draw [black] (3) to (9);
\draw [black] (4) to (7);
\tkzDefPoint(-66.50112,70.61587){A}
\tkzDefPoint(66.50112,70.61587){B}
\draw[<-,line width=0.9mm, red](A) to (B);
\tkzDefPoint(70.61587,66.50112){A}
\tkzDefPoint(70.61587,-66.50112){B}
\draw[<-,line width=0.9mm, blue](A) to (B);
\tkzDefPoint(66.50112,-70.61587){A}
\tkzDefPoint(-66.50112,-70.61587){B}
\draw[->,line width=0.9mm, red](A) to (B);
\tkzDefPoint(-70.61587,-66.50112){A}
\tkzDefPoint(-70.61587,66.50112){B}
\draw[->,line width=0.9mm, blue](A) to (B);
\node [black,circle,draw,fill=white,scale=0.75,line width=1mm] (6) at (-70.71068,-70.71068) {};
\node [black,circle,draw,fill=white,scale=0.75,line width=1mm] (10) at (-70.71068,70.71068) {};
\node [black,circle,draw,fill=white,scale=0.75,line width=1mm] (11) at (70.71068,70.71068) {};
\node [black,circle,draw,fill=white,scale=0.75,line width=1mm] (12) at (70.71068,-70.71068) {};
\end{tikzpicture}
          }
	\caption{ A map of $K_4$ on the torus with two pairs of nc-cycles: once red-blue and once green purple. The map is cut open along the blue and the red nc-cycle and forms
        one connected region. Cutting it open along the green and the purple nc-cycle would result in a disconnected surface.}
	\label{fig:cuttorus}
\end{figure}
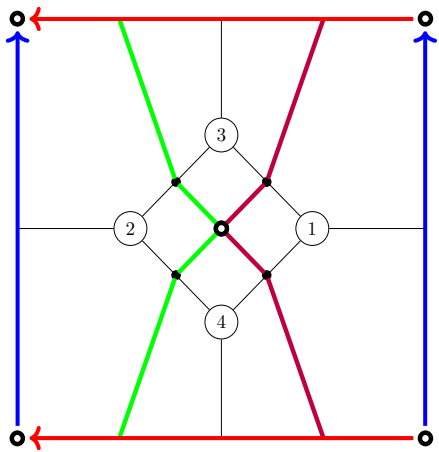

The basic strategy is to recursively search for $2g$ cycles $C_1,\dots ,C_{2g}$ that are not contractible and share one point of the surface -- by default a vertex inside a face.
These cycles are inserted into the graph by adding one vertex inside a face and afterwards new vertices on the edges and connecting these vertices to form a cycle.
Nevertheless not all such sets of nc-cycles will give a fundamental polygon when cutting the surface along them, but only those where after cutting it, the surface
forms one connected region.  Figure~\ref{fig:cuttorus} shows a map of $K_4$ on the torus with once a pair of nc-cycles along which the surface can be cut (and was cut for the drawing)
and once a pair of nc-cycles that would lead to a disconnected surface. So each time a new nc-cycle $C_i$ is found, it is checked whether the surface would still be connected after
cutting along $C_1,\dots ,C_i$ and only in this case the next cycle $C_{i+1}$ is searched (or the search is finished if $i=2g$). Otherwise a new candidate for $C_i$ is searched for.
First a heuristic determines a face for a first attempt to place the common vertex of the nc-cycles. Then
the cycles are constructed in non-decreasing length: for a fixed face $f$ in which the center is placed, it is first searched for a cycle $C_1$ crossing only one edge, then two, \dots
When constructing $C_{i+1}$, the starting length is the one of $C_i$. Though possible by adding some extra criteria determining the order, in case of cycles with the same length,
the algorithm does not prevent the same set of nc-cycles to be constructed more than once, as often already the first drawing is sufficiently good.

The result after cutting along  $C_1,\dots ,C_{2g}$ is a plane map with a fixed outer face -- the fundamental polygon. This is still a combinatorial map without
coordinates describing how to draw the graph, but  in \cite{Olaf_graphzeichnen}, Olaf Delgado~Friedrichs described a very efficient algorithm for the construction
of Schlegel diagrams in \cite{BrDeDrHa97}. In short it can be described as inserting new vertices and edges in faces to make the graph 3-connected -- a prerequisite for
an algorithm described by Tutte  \cite{Tutte_draw}.
This algorithm of Tutte is the first step to obtain coordinates: the outer face is fixed as a regular convex polygon and every vertex not on the outer
face is placed in the center of gravity of its neighbours. This first step is done by solving a system of linear equations. The result is a
drawing without crossing edges, but unfortunately often with regions with lots of vertices at very small distances on one hand and almost empty regions on the
other. To solve this problem, finally a spring embedder where the replacement is based on face areas -- see \cite{Olaf_graphzeichnen} for details -- is
applied. This very fast computation of coordinates is necessary as we will see that this coordinate computing function is sometimes called very often for a
single map.

Having found the cycles $C_1,\dots ,C_{2g}$  we are sure to find a drawing in a fundamental polygon. At this
point of the algorithm the only measure for the {\em quality} of the drawing is the minimum distance of vertices to other vertices and to edges not incident with the vertex.
If this distance is considered {\em sufficiently good} -- that is: the distances are above a prescribed minimum -- 
the default is that the computation on this input map is finished and the drawing is written to stdout.
Otherwise this minimum distance is compared to the largest minimum distance of an earlier drawing of the same graph and if larger, the formerly best drawing is replaced by the new one.

If a drawing is considered insufficient, a new attempt to find a good way to cut the graph is started. Each pair $(f,e)$, with $f$ a face and $e$ an edge in the boundary of $f$,
is a possible starting configuration and for each starting configuration only the first drawing found as described above is produced. This procedure is greedy:
once a short nc-cycle for a starting configuration is found, not even equally short cycles for this position are tested. Furthermore a short first cycle can imply superficially long
cycles with a larger index.

In order to increase the chance of finding a drawing that is considered {\em sufficiently good} early, the starting configurations $(f,e)$ are not tried in a
random order but a heuristic chooses the next configuration to be tried inside a face of maximum size where not all face-edge pairs have been tried as a
starting configuration yet.

\begin{figure}[p]
	\centering
        \resizebox{0.65\textwidth}{0.65\textwidth}
        {
          \begin{tikzpicture}[scale=0.065]
\def\vertexscale{1.00}
\def\labelscale{1.00}
\node [circle,black,draw,scale=\vertexscale] (1) at (51.59053,-36.44623) {1};
\node [circle,black,draw,scale=\vertexscale] (2) at (-51.59053,36.44623) {2};
\node [circle,black,draw,scale=\vertexscale] (3) at (72.27853,-28.11546) {3};
\node [circle,black,draw,scale=\vertexscale] (4) at (29.88427,-17.85876) {4};
\node [circle,black,draw,scale=\vertexscale] (5) at (-29.88427,17.85876) {5};
\node [circle,black,draw,scale=\vertexscale] (6) at (-72.27853,28.11546) {6};
\node [circle,black,draw,scale=\vertexscale] (7) at (-31.22801,70.98927) {7};
\node [circle,black,draw,scale=\vertexscale] (8) at (-70.98927,-31.22801) {8};
\node [circle,black,draw,scale=\vertexscale] (9) at (33.75942,8.50332) {9};
\node [circle,black,draw,scale=0.9*\vertexscale] (10) at (8.50332,-33.75942) {10};
\node [circle,black,draw,scale=0.9*\vertexscale] (11) at (-8.50332,33.75942) {11};
\node [circle,black,draw,scale=0.9*\vertexscale] (12) at (-33.75942,-8.50332) {12};
\node [circle,black,draw,scale=0.9*\vertexscale] (13) at (70.98927,31.22801) {13};
\node [circle,black,draw,scale=0.9*\vertexscale] (14) at (31.22801,-70.98927) {14};
\node [circle,black,draw,scale=0.9*\vertexscale] (15) at (-10.70864,62.25139) {15};
\node [circle,black,draw,scale=0.9*\vertexscale] (16) at (-28.11546,-72.27853) {16};
\node [circle,black,draw,scale=0.9*\vertexscale] (17) at (-62.25139,-10.70864) {17};
\node [circle,black,draw,scale=0.9*\vertexscale] (18) at (28.11546,72.27853) {18};
\node [circle,black,draw,scale=0.9*\vertexscale] (19) at (17.85876,29.88427) {19};
\node [circle,black,draw,scale=0.9*\vertexscale] (20) at (62.25139,10.70864) {20};
\node [circle,black,draw,scale=0.9*\vertexscale] (21) at (-17.85876,-29.88427) {21};
\node [circle,black,draw,scale=0.9*\vertexscale] (22) at (10.70864,-62.25139) {22};
\node [circle,black,draw,scale=0.9*\vertexscale] (23) at (-36.44623,-51.59053) {23};
\node [circle,black,draw,scale=0.9*\vertexscale] (24) at (36.44623,51.59053) {24};
\tkzDefPoint(-55.55702,83.14696){25}
\tkzDefPoint(-38.26834,-92.38795){26}
\tkzDefPoint(-70.71068,70.71068){27}
\tkzDefPoint(-83.14696,55.55702){28}
\tkzDefPoint(-19.50903,-98.07853){29}
\tkzDefPoint(0.00000,-100.00000){30}
\tkzDefPoint(19.50903,-98.07853){31}
\tkzDefPoint(-83.14696,-55.55702){32}
\tkzDefPoint(-70.71068,-70.71068){33}
\tkzDefPoint(-55.55702,-83.14696){34}
\tkzDefPoint(-98.07853,-19.50903){35}
\tkzDefPoint(-100.00000,0.00000){36}
\tkzDefPoint(-98.07853,19.50903){37}
\tkzDefPoint(83.14696,-55.55702){38}
\tkzDefPoint(70.71068,-70.71068){39}
\tkzDefPoint(55.55702,-83.14696){40}
\tkzDefPoint(98.07853,19.50903){41}
\tkzDefPoint(100.00000,0.00000){42}
\tkzDefPoint(98.07853,-19.50903){43}
\tkzDefPoint(55.55702,83.14696){44}
\tkzDefPoint(70.71068,70.71068){45}
\tkzDefPoint(83.14696,55.55702){46}
\tkzDefPoint(-19.50903,98.07853){47}
\tkzDefPoint(-0.00000,100.00000){48}
\tkzDefPoint(19.50903,98.07853){49}
\tkzDefPoint(-38.26834,92.38795){50}
\tkzDefPoint(92.38795,-38.26834){51}
\tkzDefPoint(-92.38795,-38.26834){52}
\tkzDefPoint(38.26834,92.38795){53}
\tkzDefPoint(38.26834,-92.38795){54}
\tkzDefPoint(-92.38795,38.26834){55}
\tkzDefPoint(92.38795,38.26834){56}
\draw [black] (1) to (39);
\node [draw=none,fill=none,scale=\labelscale] () at (74.24621,-74.24621) {2};
\draw [black] (1) to (4);
\draw [black] (1) to (3);
\draw [black] (2) to (27);
\node [draw=none,fill=none,scale=\labelscale] () at (-74.24621,74.24621) {1};
\draw [black] (2) to (5);
\draw [black] (2) to (6);
\draw [black] (3) to (43);
\node [draw=none,fill=none,scale=\labelscale] () at (102.98245,-20.48448) {8};
\draw [black] (3) to (38);
\node [draw=none,fill=none,scale=\labelscale] () at (87.30431,-58.33487) {7};
\draw [black] (4) to (10);
\draw [black] (4) to (9);
\draw [black] (5) to (11);
\draw [black] (5) to (12);
\draw [black] (6) to (37);
\node [draw=none,fill=none,scale=\labelscale] () at (-102.98245,20.48448) {13};
\draw [black] (6) to (28);
\node [draw=none,fill=none,scale=\labelscale] () at (-87.30431,58.33487) {14};
\draw [black] (7) to (15);
\draw [black] (7) to (25);
\node [draw=none,fill=none,scale=\labelscale] () at (-58.33487,87.30431) {3};
\draw [black] (7) to (47);
\node [draw=none,fill=none,scale=\labelscale] () at (-20.48448,102.98245) {16};
\draw [black] (8) to (32);
\node [draw=none,fill=none,scale=\labelscale] () at (-87.30431,-58.33487) {18};
\draw [black] (8) to (35);
\node [draw=none,fill=none,scale=\labelscale] () at (-102.98245,-20.48448) {3};
\draw [black] (8) to (17);
\draw [black] (9) to (19);
\draw [black] (9) to (20);
\draw [black] (10) to (21);
\draw [black] (10) to (22);
\draw [black] (11) to (15);
\draw [black] (11) to (19);
\draw [black] (12) to (17);
\draw [black] (12) to (21);
\draw [black] (13) to (46);
\node [draw=none,fill=none,scale=\labelscale] () at (87.30431,58.33487) {16};
\draw [black] (13) to (41);
\node [draw=none,fill=none,scale=\labelscale] () at (102.98245,20.48448) {6};
\draw [black] (13) to (20);
\draw [black] (14) to (40);
\node [draw=none,fill=none,scale=\labelscale] () at (58.33487,-87.30431) {6};
\draw [black] (14) to (31);
\node [draw=none,fill=none,scale=\labelscale] () at (20.48448,-102.98245) {18};
\draw [black] (14) to (22);
\draw [black] (15) to (48);
\node [draw=none,fill=none,scale=\labelscale] () at (-0.00000,105.00000) {22};
\draw [black] (16) to (29);
\node [draw=none,fill=none,scale=\labelscale] () at (-20.48448,-102.98245) {7};
\draw [black] (16) to (34);
\node [draw=none,fill=none,scale=\labelscale] () at (-58.33487,-87.30431) {13};
\draw [black] (16) to (23);
\draw [black] (17) to (36);
\node [draw=none,fill=none,scale=\labelscale] () at (-105.00000,0.00000) {20};
\draw [black] (18) to (49);
\node [draw=none,fill=none,scale=\labelscale] () at (20.48448,102.98245) {14};
\draw [black] (18) to (44);
\node [draw=none,fill=none,scale=\labelscale] () at (58.33487,87.30431) {8};
\draw [black] (18) to (24);
\draw [black] (19) to (24);
\draw [black] (20) to (42);
\node [draw=none,fill=none,scale=\labelscale] () at (105.00000,0.00000) {17};
\draw [black] (21) to (23);
\draw [black] (22) to (30);
\node [draw=none,fill=none,scale=\labelscale] () at (0.00000,-105.00000) {15};
\draw [black] (23) to (33);
\node [draw=none,fill=none,scale=\labelscale] () at (-74.24621,-74.24621) {24};
\draw [black] (24) to (45);
\node [draw=none,fill=none,scale=\labelscale] () at (74.24621,74.24621) {23};
\tkzDefPoint(-35.47990,93.49426){A}
\tkzDefPoint(35.47990,93.49426){B}
\tkzDefPoint(0.0,0.0){C}
\tkzDrawArc[<-,line width=0.9mm, red](C,B)(A)
\tkzDefPoint(41.02235,91.19850){A}
\tkzDefPoint(91.19850,41.02235){B}
\tkzDefPoint(0.0,0.0){C}
\tkzDrawArc[<-,line width=0.9mm, green](C,B)(A)
\tkzDefPoint(93.49426,35.47990){A}
\tkzDefPoint(93.49426,-35.47990){B}
\tkzDefPoint(0.0,0.0){C}
\tkzDrawArc[<-,line width=0.9mm, orange](C,B)(A)
\tkzDefPoint(91.19850,-41.02235){A}
\tkzDefPoint(41.02235,-91.19850){B}
\tkzDefPoint(0.0,0.0){C}
\tkzDrawArc[<-,line width=0.9mm, blue](C,B)(A)
\tkzDefPoint(35.47990,-93.49426){A}
\tkzDefPoint(-35.47990,-93.49426){B}
\tkzDefPoint(0.0,0.0){C}
\tkzDrawArc[->,line width=0.9mm, red](C,B)(A)
\tkzDefPoint(-41.02235,-91.19850){A}
\tkzDefPoint(-91.19850,-41.02235){B}
\tkzDefPoint(0.0,0.0){C}
\tkzDrawArc[->,line width=0.9mm, green](C,B)(A)
\tkzDefPoint(-93.49426,-35.47990){A}
\tkzDefPoint(-93.49426,35.47990){B}
\tkzDefPoint(0.0,0.0){C}
\tkzDrawArc[->,line width=0.9mm, orange](C,B)(A)
\tkzDefPoint(-91.19850,41.02235){A}
\tkzDefPoint(-41.02235,91.19850){B}
\tkzDefPoint(0.0,0.0){C}
\tkzDrawArc[->,line width=0.9mm, blue](C,B)(A)
\node [black,circle,draw,fill=white,scale=0.75,line width=1mm] (26) at (-38.26834,-92.38795) {};
\node [black,circle,draw,fill=white,scale=0.75,line width=1mm] (50) at (-38.26834,92.38795) {};
\node [black,circle,draw,fill=white,scale=0.75,line width=1mm] (51) at (92.38795,-38.26834) {};
\node [black,circle,draw,fill=white,scale=0.75,line width=1mm] (52) at (-92.38795,-38.26834) {};
\node [black,circle,draw,fill=white,scale=0.75,line width=1mm] (53) at (38.26834,92.38795) {};
\node [black,circle,draw,fill=white,scale=0.75,line width=1mm] (54) at (38.26834,-92.38795) {};
\node [black,circle,draw,fill=white,scale=0.75,line width=1mm] (55) at (-92.38795,38.26834) {};
\node [black,circle,draw,fill=white,scale=0.75,line width=1mm] (56) at (92.38795,38.26834) {};
\draw[color=black] (0.0,125.00000) node [font={\Large}] {$|V|=24, \quad   |E|=36,    \quad  |F|=10,    \quad  \mbox{genus } 2$};
\end{tikzpicture}
          }
	\caption{A minimum genus drawing of the (3,7)-cage -- the McGee graph.}
	\label{fig:mcgee}
\end{figure}
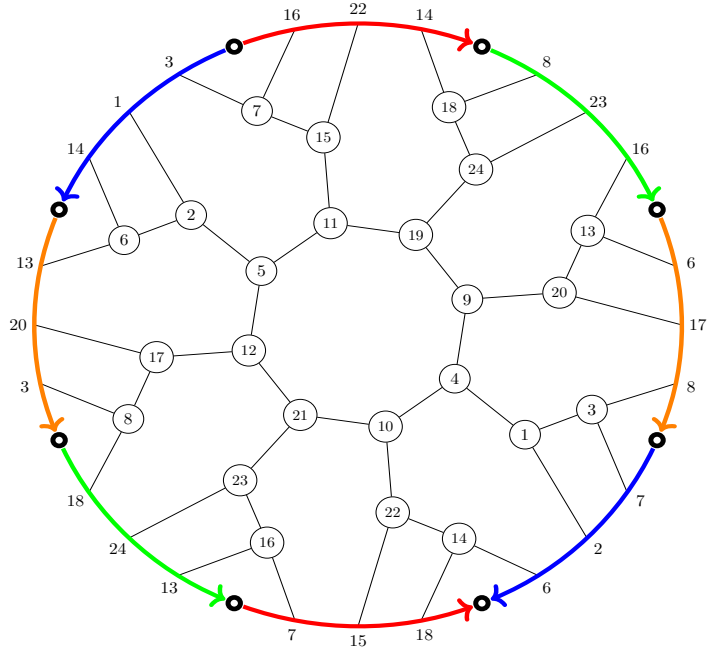

If all starting positions have been tried without a drawing being considered sufficiently good, the best -- or maybe one should say {\em least insufficient} --
drawing is written from the buffer to stdout. An example drawing of a minimum genus map of the McGee graph produced in the way described is given in
Figure~\ref{fig:mcgee}.  The points where the edges cross the fundamental polygon are labeled with the vertex number that the edges finally lead to, so that
the rotational order around each vertex can be easily read off the drawing. The symmetry of the drawing is accidental and not enforced by the drawing
algorithm. The fundamental polygon has the minimum number of sides, but is not the standard fundamental polygon for genus $2$.

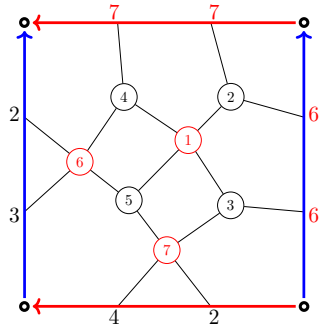
\begin{figure}[p]
	\centering
        \resizebox{0.3\textwidth}{0.3\textwidth}
        {
          \begin{tikzpicture}[scale=0.065]
\def\vertexscale{1.40}
\def\labelscale{2.00}
\node [circle,red,draw,scale=\vertexscale] (1) at (12.13209,12.13209) {1};
\node [circle,black,draw,scale=\vertexscale] (2) at (33.72393,33.72393) {2};
\node [circle,black,draw,scale=\vertexscale] (3) at (33.62343,-20.33015) {3};
\node [circle,black,draw,scale=\vertexscale] (4) at (-20.33015,33.62343) {4};
\node [circle,black,draw,scale=\vertexscale] (5) at (-17.82637,-17.82637) {5};
\node [circle,red,draw,scale=\vertexscale] (6) at (-42.64236,1.35157) {6};
\node [circle,red,draw,scale=\vertexscale] (7) at (1.35157,-42.64236) {7};
\tkzDefPoint(-70.71068,23.57023){8}
\tkzDefPoint(-70.71068,-70.71068){9}
\tkzDefPoint(-70.71068,-23.57023){10}
\tkzDefPoint(-23.57023,-70.71068){11}
\tkzDefPoint(23.57023,-70.71068){12}
\tkzDefPoint(70.71068,23.57023){13}
\tkzDefPoint(70.71068,-23.57023){14}
\tkzDefPoint(-23.57023,70.71068){15}
\tkzDefPoint(23.57023,70.71068){16}
\tkzDefPoint(-70.71068,70.71068){17}
\tkzDefPoint(70.71068,70.71068){18}
\tkzDefPoint(70.71068,-70.71068){19}
\draw [black] (1) to (2);
\draw [black] (1) to (3);
\draw [black] (1) to (5);
\draw [black] (1) to (4);
\draw [black] (2) to (16);
\node [draw=none,red,fill=none,scale=\labelscale] () at (25.24371,75.73114) {7};
\draw [black] (2) to (13);
\node [draw=none,red,fill=none,scale=\labelscale] () at (75.73114,25.24371) {6};
\draw [black] (3) to (14);
\node [draw=none,red,fill=none,scale=\labelscale] () at (75.73114,-25.24371) {6};
\draw [black] (3) to (7);
\draw [black] (4) to (15);
\node [draw=none,red,fill=none,scale=\labelscale] () at (-25.24371,75.73114) {7};
\draw [black] (4) to (6);
\draw [black] (5) to (6);
\draw [black] (5) to (7);
\draw [black] (6) to (8);
\node [draw=none,fill=none,scale=\labelscale] () at (-75.73114,25.24371) {2};
\draw [black] (6) to (10);
\node [draw=none,fill=none,scale=\labelscale] () at (-75.73114,-25.24371) {3};
\draw [black] (7) to (12);
\node [draw=none,fill=none,scale=\labelscale] () at (25.24371,-75.73114) {2};
\draw [black] (7) to (11);
\node [draw=none,fill=none,scale=\labelscale] () at (-25.24371,-75.73114) {4};
\tkzDefPoint(-66.50112,70.61587){A}
\tkzDefPoint(66.50112,70.61587){B}
\draw[<-,line width=0.9mm, red](A) to (B);
\tkzDefPoint(70.61587,66.50112){A}
\tkzDefPoint(70.61587,-66.50112){B}
\draw[<-,line width=0.9mm, blue](A) to (B);
\tkzDefPoint(66.50112,-70.61587){A}
\tkzDefPoint(-66.50112,-70.61587){B}
\draw[->,line width=0.9mm, red](A) to (B);
\tkzDefPoint(-70.61587,-66.50112){A}
\tkzDefPoint(-70.61587,66.50112){B}
\draw[->,line width=0.9mm, blue](A) to (B);
\node [black,circle,draw,fill=white,scale=0.75,line width=1mm] (9) at (-70.71068,-70.71068) {};
\node [black,circle,draw,fill=white,scale=0.75,line width=1mm] (17) at (-70.71068,70.71068) {};
\node [black,circle,draw,fill=white,scale=0.75,line width=1mm] (18) at (70.71068,70.71068) {};
\node [black,circle,draw,fill=white,scale=0.75,line width=1mm] (19) at (70.71068,-70.71068) {};
\end{tikzpicture}
          }
	\caption{A minimum genus drawing of $K_{3,4}$ forcing the face on the left of the oriented edge $[7,2]$ to be the face with the central vertex (option {\em cf 7 2}), colouring vertices of degree $4$ red (option {\em d 4 red}), making the cut cycles straight lines (option {\em s}),  and manually changing the values of {\em vertexscale} and {\em labelscale} in the header of the tikz-file from $1.0$ to $1.4$, resp.\ to $2.0$ to take the small size of the drawing into account.}
        	\label{fig:k34}
\end{figure}

The program allows several options to influence the drawings. The most important ones are:

\begin{description}
\item[if] Try to find drawings with many faces that are not cut by the nc-cycles. No complete search is done, but the option can nevertheless cause a serious increase in computing time, as after having found a {\em sufficiently good} drawing,
  it is still tried to find one with more interior faces and also new ones with the same number of interior faces are compared.
\item[cf x y] Take the face on the left of the oriented edge $[x,y]$ as the one containing the common vertex of the cutting cycles.
\item[NE x y] Do not cut through edge $\{x,y\}$ if present.
\item[b] Do not use colours for the edges of the fundamental polygon, but label them A,B,\dots . For large genera this is done automatically as too many colours are hard to distinguish.
\item[d x colourname] Colour vertices of degree $x$ with colour {\em colourname} (see Figure~\ref{fig:k34}).
\item[i] Also write the number of edges, vertices, and faces as well as the genus to the drawing (see Figure~\ref{fig:mcgee}).
\item[s] Use straight line segments for the sides of the fundamental polygon instead of cycle segments (see Figure~\ref{fig:k34}).
\end{description}

An example of a drawing produced with some of these options is given in Figure~\ref{fig:k34}. While option {\em cf} normally speeds up the program, as fewer
starting configurations have to be tested, the option {\bf NE} -- especially if used for more than one edge -- can increase the running time and even make it impossible to find a set of
cut cycles.

\subsection{Alternative centers for the nc-cycles}\label{sec:alternative}

Though planar\_draw prefers to choose the center of a face as the common center of the nc-cycles, in some special cases it might be
preferable to choose the center of an edge or a vertex instead. This can reduce the number of edge crossings with 
the boundary of the fundamental polygon, but comes with a price:

In case the common point is a center of an edge, for each vertex the order of neighbours can still be easily read off the drawing, but the edge with the cycle center is not as easily followed across the border,
which is a small disadvantage and not worse than edges crossing the border several times.
The program can be told to choose the center of an edge of the original graph as common point of the cutting cycles by using option {\em e}.

In case the common point is a vertex, this vertex will occur $4g$ times in the boundary of the fundamental polygon, so the price for possibly reducing the
number of times edges cross the boundary is that for the center vertex the rotational order of the neighbours can not that easily be read off the
drawing. This choice of common point is enforced by option {\em v}.

When the total number of crossings is very high, it can sometimes be reduced by allowing to also cut through vertices. Option {\em V} allows to cut through vertices, but each vertex is cut at most once
and the nc-cycles never use edges of the graph. This option can be combined with the option {\em NV x}, implying that vertex {\em x} must not be cut. Though implemented, option {\em V} is not recommended.

Option {\em v} can be combined with option {\em cv}:
\begin{description}
\item[cv x] Take vertex $x$ as the common point of the nc-cycles.
\end{description}

Option {\em e} can also be combined with option {\em ce}:
\begin{description}
\item[ce x y] Take the midpoint of the edge $\{x,y\}$ as the common point of the nc-cycles.
\end{description}

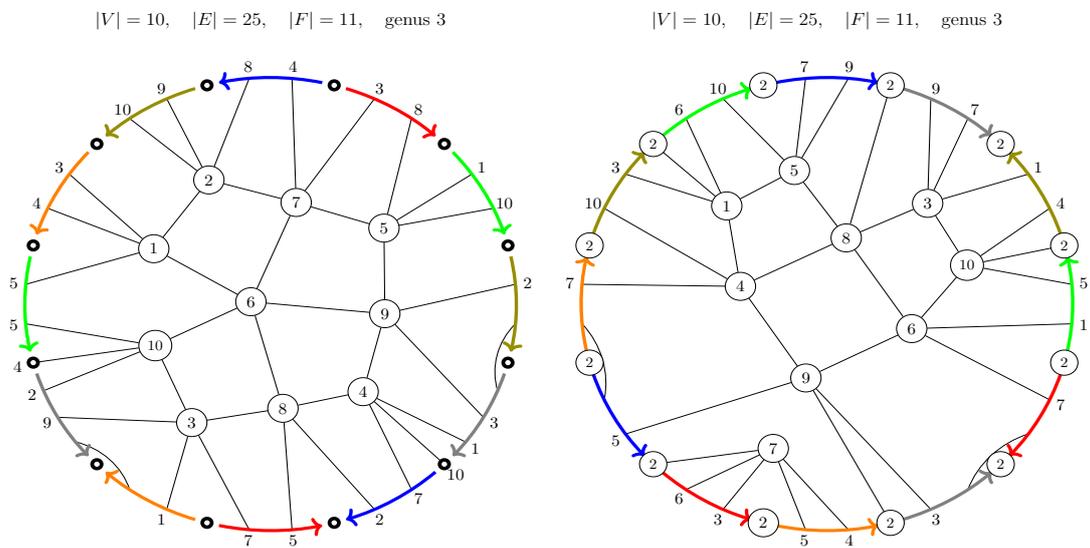
\begin{figure}[p]
	\centering
        \resizebox{0.49\textwidth}{0.49\textwidth}
        {
          \begin{tikzpicture}[scale=0.065]
\def\vertexscale{1.30}
\def\labelscale{1.30}
\node [circle,black,draw,scale=\vertexscale] (1) at (-47.55036,24.43583) {1};
\node [circle,black,draw,scale=\vertexscale] (2) at (-25.11390,54.80396) {2};
\node [circle,black,draw,scale=\vertexscale] (3) at (-32.07762,-52.33502) {3};
\node [circle,black,draw,scale=\vertexscale] (4) at (37.66395,-38.69004) {4};
\node [circle,black,draw,scale=\vertexscale] (5) at (46.14899,33.54602) {5};
\node [circle,black,draw,scale=\vertexscale] (6) at (-8.01451,1.02810) {6};
\node [circle,black,draw,scale=\vertexscale] (7) at (10.39860,44.85886) {7};
\node [circle,black,draw,scale=\vertexscale] (8) at (5.01527,-46.04090) {8};
\node [circle,black,draw,scale=\vertexscale] (9) at (46.48742,-4.26371) {9};
\node [circle,black,draw,scale=0.9*\vertexscale] (10) at (-46.93374,-18.33105) {10};
\tkzDefPoint(-25.88190,-96.59258){11}
\tkzDefPoint(8.71557,99.61947){12}
\tkzDefPoint(-8.71557,99.61947){13}
\tkzDefPoint(-8.71557,-99.61947){14}
\tkzDefPoint(8.71557,-99.61947){15}
\tkzDefPoint(-99.61947,8.71557){16}
\tkzDefPoint(-99.61947,-8.71557){17}
\tkzDefPoint(-90.63078,42.26183){18}
\tkzDefPoint(-81.91520,57.35764){19}
\tkzDefPoint(-42.26183,90.63078){20}
\tkzDefPoint(-57.35764,81.91520){21}
\tkzDefPoint(-79.33533,-60.87614){22}
\tkzDefPoint(-86.60254,-50.00000){23}
\tkzDefPoint(-92.38795,-38.26834){24}
\tkzDefPoint(57.35764,-81.91520){25}
\tkzDefPoint(42.26183,-90.63078){26}
\tkzDefPoint(99.61947,8.71557){27}
\tkzDefPoint(99.61947,-8.71557){28}
\tkzDefPoint(-42.26183,-90.63078){29}
\tkzDefPoint(-57.35764,-81.91520){30}
\tkzDefPoint(81.91520,57.35764){31}
\tkzDefPoint(90.63078,42.26183){32}
\tkzDefPoint(92.38795,-38.26834){33}
\tkzDefPoint(86.60254,-50.00000){34}
\tkzDefPoint(79.33533,-60.87614){35}
\tkzDefPoint(42.26183,90.63078){36}
\tkzDefPoint(57.35764,81.91520){37}
\tkzDefPoint(25.88190,96.59258){38}
\tkzDefPoint(70.71068,-70.71068){39}
\tkzDefPoint(-70.71068,-70.71068){40}
\tkzDefPoint(-96.59258,25.88190){41}
\tkzDefPoint(70.71068,70.71068){42}
\tkzDefPoint(25.88190,-96.59258){43}
\tkzDefPoint(-25.88190,96.59258){44}
\tkzDefPoint(96.59258,25.88190){45}
\tkzDefPoint(-96.59258,-25.88190){46}
\tkzDefPoint(96.59258,-25.88190){47}
\tkzDefPoint(-70.71068,70.71068){48}
\draw [black] (1) to (2);
\draw [black] (1) to (6);
\draw [black] (1) to (16);
\node [draw=none,fill=none,scale=\labelscale] () at (-104.60044,9.15135) {5};
\draw [black] (1) to (18);
\node [draw=none,fill=none,scale=\labelscale] () at (-95.16232,44.37492) {4};
\draw [black] (1) to (19);
\node [draw=none,fill=none,scale=\labelscale] () at (-86.01096,60.22553) {3};
\draw [black] (2) to (21);
\node [draw=none,fill=none,scale=\labelscale] () at (-60.22553,86.01096) {10};
\draw [black] (2) to (20);
\node [draw=none,fill=none,scale=\labelscale] () at (-44.37492,95.16232) {9};
\draw [black] (2) to (13);
\node [draw=none,fill=none,scale=\labelscale] () at (-9.15135,104.60044) {8};
\draw [black] (2) to (7);
\draw [black] (3) to (29);
\node [draw=none,fill=none,scale=\labelscale] () at (-44.37492,-95.16232) {1};
\draw [black] (3) to (23);
\node [draw=none,fill=none,scale=\labelscale] () at (-90.93267,-52.50000) {9};
\draw [black] (3) to (10);
\draw [black] (3) to (8);
\draw [black] (3) to (14);
\node [draw=none,fill=none,scale=\labelscale] () at (-9.15135,-104.60044) {7};
\draw [black] (4) to (35);
\node [draw=none,fill=none,scale=\labelscale] () at (83.30210,-63.91995) {1};
\draw [black] (4) to (39);
\draw [black] (4) to (25);
\node [draw=none,fill=none,scale=\labelscale] () at (60.22553,-86.01096) {7};
\draw [black] (4) to (8);
\draw [black] (4) to (9);
\draw [black] (5) to (7);
\draw [black] (5) to (37);
\node [draw=none,fill=none,scale=\labelscale] () at (60.22553,86.01096) {8};
\draw [black] (5) to (31);
\node [draw=none,fill=none,scale=\labelscale] () at (86.01096,60.22553) {1};
\draw [black] (5) to (32);
\node [draw=none,fill=none,scale=\labelscale] () at (95.16232,44.37492) {10};
\draw [black] (5) to (9);
\draw [black] (6) to (8);
\draw [black] (6) to (10);
\draw [black] (6) to (7);
\draw [black] (6) to (9);
\draw [black] (7) to (12);
\node [draw=none,fill=none,scale=\labelscale] () at (9.15135,104.60044) {4};
\draw [black] (7) to (36);
\node [draw=none,fill=none,scale=\labelscale] () at (44.37492,95.16232) {3};
\draw [black] (8) to (15);
\node [draw=none,fill=none,scale=\labelscale] () at (9.15135,-104.60044) {5};
\draw [black] (8) to (26);
\node [draw=none,fill=none,scale=\labelscale] () at (44.37492,-95.16232) {2};
\draw [black] (9) to (27);
\node [draw=none,fill=none,scale=\labelscale] () at (104.60044,9.15135) {2};
\draw [black] (9) to (34);
\node [draw=none,fill=none,scale=\labelscale] () at (90.93267,-52.50000) {3};
\draw [black] (10) to (46);
\draw [black] (10) to (17);
\node [draw=none,fill=none,scale=\labelscale] () at (-104.60044,-9.15135) {5};
\draw [black] (10) to (24);
\node [draw=none,fill=none,scale=\labelscale] () at (-97.00735,-40.18176) {2};
\tkzDefPoint(-79.33533,-60.87614){A}
\tkzDefPoint(-65.61263,-68.53985){B}
\tkzDefPoint(-57.35764,-81.91520){C}
\tkzCircumCenter(A,B,C)\tkzGetPoint{D}
\tkzDrawArc[black](D,C)(A)
\tkzDefPoint(99.61947,-8.71557){A}
\tkzDefPoint(92.16356,-22.55228){B}
\tkzDefPoint(92.38795,-38.26834){C}
\tkzCircumCenter(A,B,C)\tkzGetPoint{D}
\tkzDrawArc[black](D,A)(C)
\tkzDefPoint(67.08825,-74.15637){A}
\tkzDefPoint(30.67718,-95.17831){B}
\tkzDefPoint(0.0,0.0){C}
\tkzDrawArc[<-,line width=0.9mm, blue](C,B)(A)
\tkzDefPoint(21.02194,-97.76542){A}
\tkzDefPoint(-21.02194,-97.76542){B}
\tkzDefPoint(0.0,0.0){C}
\tkzDrawArc[->,line width=0.9mm, red](C,B)(A)
\tkzDefPoint(-30.67718,-95.17831){A}
\tkzDefPoint(-67.08825,-74.15637){B}
\tkzDefPoint(0.0,0.0){C}
\tkzDrawArc[<-,line width=0.9mm, orange](C,B)(A)
\tkzDefPoint(-74.15637,-67.08825){A}
\tkzDefPoint(-95.17831,-30.67718){B}
\tkzDefPoint(0.0,0.0){C}
\tkzDrawArc[->,line width=0.9mm, gray](C,B)(A)
\tkzDefPoint(-97.76542,-21.02194){A}
\tkzDefPoint(-97.76542,21.02194){B}
\tkzDefPoint(0.0,0.0){C}
\tkzDrawArc[->,line width=0.9mm, green](C,B)(A)
\tkzDefPoint(-95.17831,30.67718){A}
\tkzDefPoint(-74.15637,67.08825){B}
\tkzDefPoint(0.0,0.0){C}
\tkzDrawArc[->,line width=0.9mm, orange](C,B)(A)
\tkzDefPoint(-67.08825,74.15637){A}
\tkzDefPoint(-30.67718,95.17831){B}
\tkzDefPoint(0.0,0.0){C}
\tkzDrawArc[->,line width=0.9mm, olive](C,B)(A)
\tkzDefPoint(-21.02194,97.76542){A}
\tkzDefPoint(21.02194,97.76542){B}
\tkzDefPoint(0.0,0.0){C}
\tkzDrawArc[->,line width=0.9mm, blue](C,B)(A)
\tkzDefPoint(30.67718,95.17831){A}
\tkzDefPoint(67.08825,74.15637){B}
\tkzDefPoint(0.0,0.0){C}
\tkzDrawArc[<-,line width=0.9mm, red](C,B)(A)
\tkzDefPoint(74.15637,67.08825){A}
\tkzDefPoint(95.17831,30.67718){B}
\tkzDefPoint(0.0,0.0){C}
\tkzDrawArc[<-,line width=0.9mm, green](C,B)(A)
\tkzDefPoint(97.76542,21.02194){A}
\tkzDefPoint(97.76542,-21.02194){B}
\tkzDefPoint(0.0,0.0){C}
\tkzDrawArc[<-,line width=0.9mm, olive](C,B)(A)
\tkzDefPoint(95.17831,-30.67718){A}
\tkzDefPoint(74.15637,-67.08825){B}
\tkzDefPoint(0.0,0.0){C}
\tkzDrawArc[<-,line width=0.9mm, gray](C,B)(A)
\node [black,circle,draw,fill=white,scale=0.75,line width=1mm] (11) at (-25.88190,-96.59258) {};
\node [black,circle,draw,fill=white,scale=0.75,line width=1mm] (38) at (25.88190,96.59258) {};
\node [black,circle,draw,fill=white,scale=0.75,line width=1mm] (39) at (70.71068,-70.71068) {};
\node [draw=none,fill=none,scale=\labelscale]  () at (75.30687,-75.30687) {10};
\node [black,circle,draw,fill=white,scale=0.75,line width=1mm] (40) at (-70.71068,-70.71068) {};
\node [black,circle,draw,fill=white,scale=0.75,line width=1mm] (41) at (-96.59258,25.88190) {};
\node [black,circle,draw,fill=white,scale=0.75,line width=1mm] (42) at (70.71068,70.71068) {};
\node [black,circle,draw,fill=white,scale=0.75,line width=1mm] (43) at (25.88190,-96.59258) {};
\node [black,circle,draw,fill=white,scale=0.75,line width=1mm] (44) at (-25.88190,96.59258) {};
\node [black,circle,draw,fill=white,scale=0.75,line width=1mm] (45) at (96.59258,25.88190) {};
\node [black,circle,draw,fill=white,scale=0.75,line width=1mm] (46) at (-96.59258,-25.88190) {};
\node [draw=none,fill=none,scale=\labelscale]  () at (-102.87110,-27.56423) {4};
\node [black,circle,draw,fill=white,scale=0.75,line width=1mm] (47) at (96.59258,-25.88190) {};
\node [black,circle,draw,fill=white,scale=0.75,line width=1mm] (48) at (-70.71068,70.71068) {};
\draw[color=black] (0.0,125.00000) node [font={\Large}] {$|V|=10, \quad   |E|=25,    \quad  |F|=11,    \quad  \mbox{genus } 3$};
\end{tikzpicture}
        }
               \resizebox{0.49\textwidth}{0.49\textwidth}
        {
          \begin{tikzpicture}[scale=0.065]
\def\vertexscale{1.30}
\def\labelscale{1.30}
\node [circle,black,draw,scale=\vertexscale] (1) at (-40.77401,43.34215) {1};
\tkzDefPoint(-25.88190,-96.59258){2}
\node [circle,black,draw,scale=\vertexscale] (3) at (40.98139,44.47633) {3};
\node [circle,black,draw,scale=\vertexscale] (4) at (-35.21290,7.80859) {4};
\node [circle,black,draw,scale=\vertexscale] (5) at (-13.22833,58.99356) {5};
\node [circle,black,draw,scale=\vertexscale] (6) at (34.50512,-10.84777) {6};
\node [circle,black,draw,scale=\vertexscale] (7) at (-21.83975,-63.73322) {7};
\node [circle,black,draw,scale=\vertexscale] (8) at (7.84600,29.22176) {8};
\node [circle,black,draw,scale=\vertexscale] (9) at (-8.55485,-32.63147) {9};
\node [circle,black,draw,scale=0.9*\vertexscale] (10) at (57.01388,17.16023) {10};
\tkzDefPoint(-90.63078,-42.26183){11}
\tkzDefPoint(-81.91520,-57.35764){12}
\tkzDefPoint(-42.26183,-90.63078){13}
\tkzDefPoint(-57.35764,-81.91520){14}
\tkzDefPoint(99.61947,8.71557){15}
\tkzDefPoint(99.61947,-8.71557){16}
\tkzDefPoint(8.71557,-99.61947){17}
\tkzDefPoint(-8.71557,-99.61947){18}
\tkzDefPoint(90.63078,42.26183){19}
\tkzDefPoint(81.91520,57.35764){20}
\tkzDefPoint(42.26183,-90.63078){21}
\tkzDefPoint(57.35764,-81.91520){22}
\tkzDefPoint(-8.71557,99.61947){23}
\tkzDefPoint(8.71557,99.61947){24}
\tkzDefPoint(90.63078,-42.26183){25}
\tkzDefPoint(81.91520,-57.35764){26}
\tkzDefPoint(-99.61947,-8.71557){27}
\tkzDefPoint(-99.61947,8.71557){28}
\tkzDefPoint(42.26183,90.63078){29}
\tkzDefPoint(57.35764,81.91520){30}
\tkzDefPoint(-57.35764,81.91520){31}
\tkzDefPoint(-42.26183,90.63078){32}
\tkzDefPoint(-90.63078,42.26183){33}
\tkzDefPoint(-81.91520,57.35764){34}
\tkzDefPoint(-96.59258,-25.88190){35}
\tkzDefPoint(-25.88190,96.59258){36}
\tkzDefPoint(96.59258,25.88190){37}
\tkzDefPoint(-96.59258,25.88190){38}
\tkzDefPoint(25.88190,-96.59258){39}
\tkzDefPoint(25.88190,96.59258){40}
\tkzDefPoint(-70.71068,-70.71068){41}
\tkzDefPoint(96.59258,-25.88190){42}
\tkzDefPoint(-70.71068,70.71068){43}
\tkzDefPoint(70.71068,70.71068){44}
\tkzDefPoint(70.71068,-70.71068){45}
\draw [black] (1) to (43);
\draw [black] (1) to (31);
\node [draw=none,fill=none,scale=\labelscale] () at (-60.22553,86.01096) {6};
\draw [black] (1) to (5);
\draw [black] (1) to (4);
\draw [black] (1) to (34);
\node [draw=none,fill=none,scale=\labelscale] () at (-86.01096,60.22553) {3};
\draw [black] (3) to (20);
\node [draw=none,fill=none,scale=\labelscale] () at (86.01096,60.22553) {1};
\draw [black] (3) to (10);
\draw [black] (3) to (8);
\draw [black] (3) to (29);
\node [draw=none,fill=none,scale=\labelscale] () at (44.37492,95.16232) {9};
\draw [black] (3) to (30);
\node [draw=none,fill=none,scale=\labelscale] () at (60.22553,86.01096) {7};
\draw [black] (4) to (8);
\draw [black] (4) to (9);
\draw [black] (4) to (28);
\node [draw=none,fill=none,scale=\labelscale] () at (-104.60044,9.15135) {7};
\draw [black] (4) to (33);
\node [draw=none,fill=none,scale=\labelscale] () at (-95.16232,44.37492) {10};
\draw [black] (5) to (23);
\node [draw=none,fill=none,scale=\labelscale] () at (-9.15135,104.60044) {7};
\draw [black] (5) to (24);
\node [draw=none,fill=none,scale=\labelscale] () at (9.15135,104.60044) {9};
\draw [black] (5) to (8);
\draw [black] (5) to (32);
\node [draw=none,fill=none,scale=\labelscale] () at (-44.37492,95.16232) {10};
\draw [black] (6) to (8);
\draw [black] (6) to (10);
\draw [black] (6) to (16);
\node [draw=none,fill=none,scale=\labelscale] () at (104.60044,-9.15135) {1};
\draw [black] (6) to (25);
\node [draw=none,fill=none,scale=\labelscale] () at (95.16232,-44.37492) {7};
\draw [black] (6) to (9);
\draw [black] (7) to (41);
\draw [black] (7) to (17);
\node [draw=none,fill=none,scale=\labelscale] () at (9.15135,-104.60044) {4};
\draw [black] (7) to (18);
\node [draw=none,fill=none,scale=\labelscale] () at (-9.15135,-104.60044) {5};
\draw [black] (7) to (13);
\node [draw=none,fill=none,scale=\labelscale] () at (-44.37492,-95.16232) {3};
\draw [black] (7) to (14);
\node [draw=none,fill=none,scale=\labelscale] () at (-60.22553,-86.01096) {6};
\draw [black] (8) to (40);
\draw [black] (9) to (21);
\node [draw=none,fill=none,scale=\labelscale] () at (44.37492,-95.16232) {3};
\draw [black] (9) to (39);
\draw [black] (9) to (12);
\node [draw=none,fill=none,scale=\labelscale] () at (-86.01096,-60.22553) {5};
\draw [black] (10) to (19);
\node [draw=none,fill=none,scale=\labelscale] () at (95.16232,44.37492) {4};
\draw [black] (10) to (37);
\draw [black] (10) to (15);
\node [draw=none,fill=none,scale=\labelscale] () at (104.60044,9.15135) {5};
\tkzDefPoint(-90.63078,-42.26183){A}
\tkzDefPoint(-91.32012,-24.46915){B}
\tkzDefPoint(-99.61947,-8.71557){C}
\tkzCircumCenter(A,B,C)\tkzGetPoint{D}
\tkzDrawArc[black](D,A)(C)
\tkzDefPoint(57.35764,-81.91520){A}
\tkzDefPoint(66.85097,-66.85097){B}
\tkzDefPoint(81.91520,-57.35764){C}
\tkzCircumCenter(A,B,C)\tkzGetPoint{D}
\tkzDrawArc[black](D,C)(A)
\tkzDefPoint(97.76542,21.02194){A}
\tkzDefPoint(97.76542,-21.02194){B}
\tkzDefPoint(0.0,0.0){C}
\tkzDrawArc[->,line width=0.9mm, green](C,B)(A)
\tkzDefPoint(95.17831,-30.67718){A}
\tkzDefPoint(74.15637,-67.08825){B}
\tkzDefPoint(0.0,0.0){C}
\tkzDrawArc[<-,line width=0.9mm, red](C,B)(A)
\tkzDefPoint(67.08825,-74.15637){A}
\tkzDefPoint(30.67718,-95.17831){B}
\tkzDefPoint(0.0,0.0){C}
\tkzDrawArc[->,line width=0.9mm, gray](C,B)(A)
\tkzDefPoint(21.02194,-97.76542){A}
\tkzDefPoint(-21.02194,-97.76542){B}
\tkzDefPoint(0.0,0.0){C}
\tkzDrawArc[->,line width=0.9mm, orange](C,B)(A)
\tkzDefPoint(-30.67718,-95.17831){A}
\tkzDefPoint(-67.08825,-74.15637){B}
\tkzDefPoint(0.0,0.0){C}
\tkzDrawArc[->,line width=0.9mm, red](C,B)(A)
\tkzDefPoint(-74.15637,-67.08825){A}
\tkzDefPoint(-95.17831,-30.67718){B}
\tkzDefPoint(0.0,0.0){C}
\tkzDrawArc[->,line width=0.9mm, blue](C,B)(A)
\tkzDefPoint(-97.76542,-21.02194){A}
\tkzDefPoint(-97.76542,21.02194){B}
\tkzDefPoint(0.0,0.0){C}
\tkzDrawArc[<-,line width=0.9mm, orange](C,B)(A)
\tkzDefPoint(-95.17831,30.67718){A}
\tkzDefPoint(-74.15637,67.08825){B}
\tkzDefPoint(0.0,0.0){C}
\tkzDrawArc[<-,line width=0.9mm, olive](C,B)(A)
\tkzDefPoint(-67.08825,74.15637){A}
\tkzDefPoint(-30.67718,95.17831){B}
\tkzDefPoint(0.0,0.0){C}
\tkzDrawArc[<-,line width=0.9mm, green](C,B)(A)
\tkzDefPoint(-21.02194,97.76542){A}
\tkzDefPoint(21.02194,97.76542){B}
\tkzDefPoint(0.0,0.0){C}
\tkzDrawArc[<-,line width=0.9mm, blue](C,B)(A)
\tkzDefPoint(30.67718,95.17831){A}
\tkzDefPoint(67.08825,74.15637){B}
\tkzDefPoint(0.0,0.0){C}
\tkzDrawArc[<-,line width=0.9mm, gray](C,B)(A)
\tkzDefPoint(74.15637,67.08825){A}
\tkzDefPoint(95.17831,30.67718){B}
\tkzDefPoint(0.0,0.0){C}
\tkzDrawArc[->,line width=0.9mm, olive](C,B)(A)
\node [circle,black,draw,fill=white,scale=\vertexscale] (2) at (-25.88190,-96.59258) {2};
\node [circle,black,draw,fill=white,scale=0.9*\vertexscale] (35) at (-96.59258,-25.88190) {2};
\node [circle,black,draw,fill=white,scale=0.9*\vertexscale] (36) at (-25.88190,96.59258) {2};
\node [circle,black,draw,fill=white,scale=0.9*\vertexscale] (37) at (96.59258,25.88190) {2};
\node [circle,black,draw,fill=white,scale=0.9*\vertexscale] (38) at (-96.59258,25.88190) {2};
\node [circle,black,draw,fill=white,scale=0.9*\vertexscale] (39) at (25.88190,-96.59258) {2};
\node [circle,black,draw,fill=white,scale=0.9*\vertexscale] (40) at (25.88190,96.59258) {2};
\node [circle,black,draw,fill=white,scale=0.9*\vertexscale] (41) at (-70.71068,-70.71068) {2};
\node [circle,black,draw,fill=white,scale=0.9*\vertexscale] (42) at (96.59258,-25.88190) {2};
\node [circle,black,draw,fill=white,scale=0.9*\vertexscale] (43) at (-70.71068,70.71068) {2};
\node [circle,black,draw,fill=white,scale=0.9*\vertexscale] (44) at (70.71068,70.71068) {2};
\node [circle,black,draw,fill=white,scale=0.9*\vertexscale] (45) at (70.71068,-70.71068) {2};
\draw[color=black] (0.0,125.00000) node [font={\Large}] {$|V|=10, \quad   |E|=25,    \quad  |F|=11,    \quad  \mbox{genus } 3$};
\end{tikzpicture}
          }
	\caption{Two different of the 167 non-isomorphic minimum genus embeddings of the $(5,4)$-cage. The first with an edge as the center of the nc-cycles and the second with a vertex as the center.}
	\label{fig:54cage}
\end{figure}

Figure~\ref{fig:54cage} shows two different of the 167 non-isomorphic minimum genus embeddings of the $(5,4)$-cage. The first drawing is with an edge as the center of the nc-cycles and the second with a vertex as the center.
       
\section{Drawings with disjoint nc-cycles}\label{sec:tubes}

\begin{figure}[p]
	\centering
        \resizebox{0.5\textwidth}{0.5\textwidth}
        {
          \begin{tikzpicture}[scale=0.07]
\def\colouredvertexscale{0.88}
\def\vertexscale{0.98}
\node [circle,black,draw,scale=\vertexscale] (1) at (-77.85114,-8.06972) {1};
\node [circle,black,draw,scale=\vertexscale] (2) at (68.41526,6.14020) {2};
\node [circle,black,draw,scale=\vertexscale] (3) at (-62.40048,6.94895) {3};
\node [circle,black,draw,scale=\vertexscale] (4) at (-97.49279,-22.25209) {4};
\node [circle,black,draw,scale=\vertexscale] (5) at (74.96315,-16.07510) {5};
\node [circle,black,draw,scale=\vertexscale] (6) at (-9.75607,-33.10146) {6};
\node [circle,black,draw,scale=\vertexscale] (7) at (57.13331,50.65016) {7};
\node [circle,black,draw,scale=\vertexscale] (8) at (-49.91696,-7.09663) {8};
\node [circle,black,draw,scale=\vertexscale] (9) at (-43.38837,-90.09689) {9};
\node [circle,black,draw,scale=0.9*\vertexscale] (10) at (-78.18315,62.34898) {10};
\node [circle,black,draw,scale=0.9*\vertexscale] (11) at (97.49279,-22.25209) {11};
\node [circle,black,draw,scale=0.9*\vertexscale] (12) at (51.91295,-20.42377) {12};
\node [circle,black,draw,scale=0.9*\vertexscale] (13) at (-17.79826,-52.21376) {13};
\node [circle,black,draw,scale=0.9*\vertexscale] (14) at (-7.65783,-5.39744) {14};
\node [circle,black,draw,scale=0.9*\vertexscale] (15) at (78.18315,62.34898) {15};
\node [circle,black,draw,scale=0.9*\vertexscale] (16) at (35.72323,26.85328) {16};
\node [circle,black,draw,scale=0.9*\vertexscale] (17) at (-41.31858,8.51657) {17};
\node [circle,black,draw,scale=0.9*\vertexscale] (18) at (-49.68333,-38.76932) {18};
\node [circle,black,draw,scale=0.9*\vertexscale] (19) at (43.38837,-90.09689) {19};
\node [circle,black,draw,scale=0.9*\vertexscale] (20) at (-32.21515,-64.73905) {20};
\node [circle,black,draw,scale=0.9*\vertexscale] (21) at (-37.68032,35.31320) {21};
\node [circle,black,draw,scale=0.9*\vertexscale] (22) at (-0.00000,100.00000) {22};
\node [circle,black,draw,scale=0.9*\vertexscale] (23) at (31.94029,-2.62775) {23};
\node [circle,black,draw,scale=0.9*\vertexscale] (24) at (18.15312,-35.07556) {24};
\node [circle,black,draw,scale=0.9*\vertexscale] (25) at (2.75881,57.14775) {25};
\node [circle,black,draw,scale=0.9*\vertexscale] (26) at (11.12722,8.17928) {26};
\node [circle,blue, thick, densely dotted,text=black,draw,scale=1.3*\colouredvertexscale] (27) at (26.92845,51.13786) {18};
\node [circle,green,text=black,draw,scale=0.9*\colouredvertexscale] (28) at (51.85642,-3.16035) {17};
\node [circle,red,text=black,draw,scale=\colouredvertexscale] (29) at (66.18407,20.41755) {1};
\node [circle,green,text=black,draw,scale=0.9*\colouredvertexscale] (30) at (46.20146,15.94756) {13};
\node [circle,red,text=black,draw,scale=\colouredvertexscale] (31) at (62.97986,32.55934) {3};
\node [circle,green,text=black,draw,scale=\colouredvertexscale] (32) at (-21.79163,-17.98823) {2};
\node [circle,green,text=black,draw,scale=\colouredvertexscale] (33) at (57.54020,9.57876) {6};
\node [circle,green,text=black,draw,scale=0.9*\colouredvertexscale] (34) at (-27.45846,-32.05083) {16};
\node [circle,green,text=black,draw,scale=0.9*\colouredvertexscale] (35) at (-32.82921,-10.53159) {12};
\node [circle,blue, thick, densely dotted,text=black,draw,scale=1.3*\colouredvertexscale] (36) at (-69.07816,-32.81959) {25};
\node [circle,red,text=black,draw,scale=\colouredvertexscale] (37) at (-71.12804,33.15891) {7};
\node [circle,red,text=black,draw,scale=\colouredvertexscale] (38) at (-76.28404,13.05258) {2};
\draw [black] (1) to (4);
\draw [black] (1) to (38);
\draw [black] (1) to (3);
\draw [black] (2) to (33);
\draw [black] (2) to (29);
\draw [black] (2) to (5);
\draw [black] (3) to (8);
\draw [black] (3) to (37);
\draw [black] (4) to (10);
\draw [black] (4) to (9);
\draw [black] (5) to (12);
\draw [black] (5) to (11);
\draw [black] (6) to (14);
\draw [black] (6) to (13);
\draw [black] (6) to (32);
\draw [black] (7) to (16);
\draw [black] (7) to (15);
\draw [black] (7) to (31);
\draw [black] (8) to (17);
\draw [black] (8) to (18);
\draw [black] (9) to (20);
\draw [black] (9) to (19);
\draw [black] (10) to (22);
\draw [black] (10) to (21);
\draw [black] (11) to (19);
\draw [black] (11) to (15);
\draw [black] (12) to (23);
\draw [black] (12) to (28);
\draw [black] (13) to (20);
\draw [black] (13) to (34);
\draw [black] (14) to (24);
\draw [black] (14) to (21);
\draw [black] (15) to (22);
\draw [black] (16) to (23);
\draw [black] (16) to (30);
\draw [black] (17) to (35);
\draw [black] (17) to (21);
\draw [black] (18) to (20);
\draw [black] (18) to (36);
\draw [black] (19) to (24);
\draw [black] (22) to (25);
\draw [black] (23) to (26);
\draw [black] (24) to (26);
\draw [black] (25) to (26);
\draw [black] (25) to (27);
\draw [green, thick, densely dotted] (28) to (30);
\draw [green, thick, densely dotted] (28) to (33);
\draw [red, bend right=20, thick, densely dotted] (29) to (31);
\draw [red, bend left=20, thick, densely dotted] (29) to (31);
\draw [green, thick, densely dotted] (30) to (33);
\draw [green, thick, densely dotted] (32) to (34);
\draw [green, thick, densely dotted] (32) to (35);
\draw [green, thick, densely dotted] (34) to (35);
\draw [red, bend right=20, thick, densely dotted] (37) to (38);
\draw [red, bend left=20, thick, densely dotted] (37) to (38);
\draw[color=black] (0.0,125.00000) node [font={\huge}] {$|V|=26, \quad   |E|=39,    \quad  |F|=9,    \quad  \mbox{genus } 3$};
\end{tikzpicture}
          }
	\caption{A drawing with disjoint nc-cycles of a cubic graph with 26 vertices and genus 3. The graph was chosen to display nc-cycles of length 1, 2, and
          3.  Dashed cycles of the same colour have to be identified (or connected by a tube). So the blue 1-cycles (that is: loops) must be identified. The
          identification gluing ends of edges to each other is unique, but in order to be able to easily read off the neighbours, inside the loop, the number of
          the vertex on the other side of the incident edge is written. The red 2-cycles and the green 3-cycles can be identified in several ways.  Here the
          labels also indicate how they must be identified. E.g.:\ the red vertex labeled $3$ says that this edge from $7$ goes to vertex $3$, so it must be
          identified with the red vertex labeled $7$ that says that this edge coming from $3$ goes to $7$. So also in these drawings the rotational order of the
          neighbours can be directly read off the drawing.\\ In drawings with disjoint nc-cycles one should be careful with the option {\em d x
            colour}. Although the cycles that have to be identified are dashed, vertices coloured due to their degree and vertices as part of coloured and
          dashed cycles could be mixed up at first sight.}
	\label{fig:exampleT}
\end{figure}
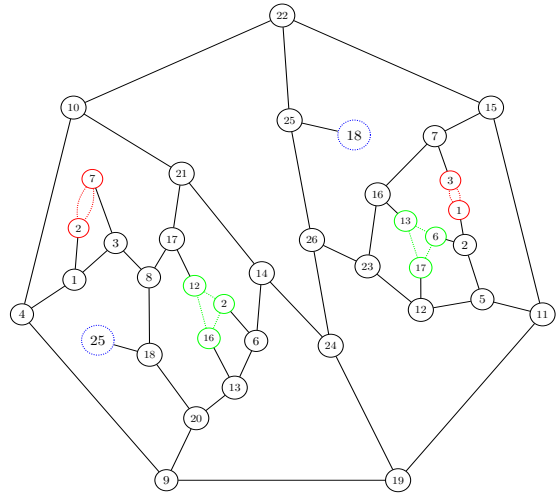

Using disjoint nc-cycles to cut the map open, we do not get a drawing in a fundamental polygon, but each nc-cycle used for cutting the map introduces two new
faces that correspond to the cycle.  In this case, a graph of genus $g$ can be transformed into a planar graph by cutting only along $g$ nc-cycles instead of
$2g$. Disjoint nc-cycles to cut the graph are computed in a similar heuristic way and by also constructing the cycles in non-decreasing order of their
lengths. Also in this case there is no complete search: for a start configuration $(f,e)$ only one shortest nc-cycle is considered as the first of $g$ cycles
that are used to cut the map. Also here the search stops as soon as a drawing is found that is considered {\em ``good enough''} with respect to the distances
of the vertices to other vertices and to edges not incident with the vertex. An example of such a drawing together with an explanation of how to interpret it is given in Figure~\ref{fig:exampleT}.

\begin{figure}[p]
	\centering
        \resizebox{0.5\textwidth}{0.5\textwidth}
        {
          \input{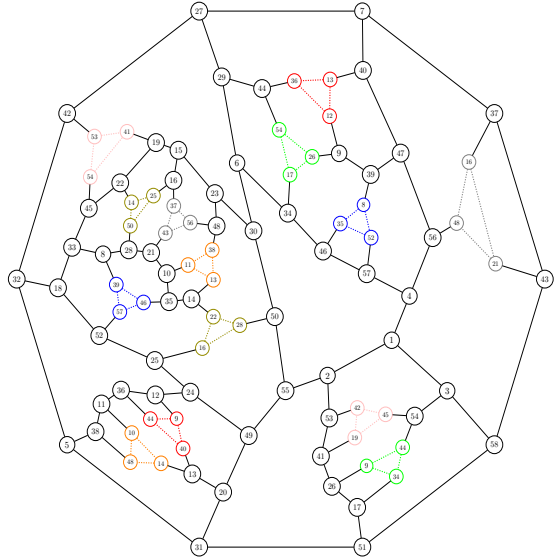}
          }
	\caption{A drawing with disjoint nc-cycles of a polyhedral map of a $(3,9)$-cage. The map has $58$ vertices and genus $7$.  As the map is polyhedral,
          each nc-cycle crosses at least three edges, so the number of $42$ new vertices in this drawing is smallest possible. As the outer face is a $10$-gon,
          $90$ vertices are inside the boundary face, while for drawings in a fundamental polygon at most $58$ vertices are inside the outer face and the
          crossings with nc-cycles are distributed over the boundary of the outer face. }
	\label{fig:9cage}
\end{figure}

Although one could argue whether drawings inside a fundamental polygon or with disjoint nc-cycles are to be preferred, when it comes to efficiency, there is
clearly an advantage for drawings inside a fundamental polygon: when drawing maps inside a fundamental polygon, all intersections are points on the boundary of
the outer face, which can be easily distributed over the circumference. Inside the outer circle there are only vertices of the graph.  When
drawing graphs with disjoint nc-cycles, we have fewer cut cycles, but each cut through an edge leads to two new vertices, which will in most cases not be on the
boundary of the outer face. So for drawings with disjoint nc-cycles, sometimes many more vertices have to fit inside the boundary of the outer cycle. This makes
vertices close to each other or even collisions of vertices much more likely and leads to more rejected drawings. An example of a case where many more vertices
have to fit inside the outer polygon is given in Figure~\ref{fig:9cage}.

Furthermore, while for drawings inside a fundamental polygon there is a unique outer face, for drawings with disjoint nc-cycles, there are as many ways to
choose the outer face as there are faces in the graph after cutting it open.  In order to limit the possibilities a bit, only faces that are at most one smaller
than the maximum face size in the uncut graph are considered (if possible) -- but in some cases also with this restriction all faces are considered.  The
combination of many more possibilities and the fact that more vertices have to fit into the boundary of the outer face, leads to much longer running times for
drawings with disjoint nc-cycles compared to drawings in fundamental polygons.

An example for the difference in efficiency: when computing drawings inside a fundamental polygon for all $741$ non-isomorphic minimum genus maps of the eighteen $(3,9)$-cages, in total $1\ 147$ drawings were
computed and tested ($3.2$ seconds). So in many cases even the first drawing produced had sufficient distances between the vertices. When computing drawings with disjoint
nc-cycles, in total $1\ 100\ 166$ drawings were produced, corresponding to $54\ 387$ different ways to cut the maps open ($2\ 840$ seconds).

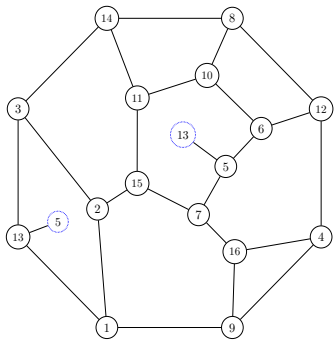
\begin{figure}[p]
	\centering
        \resizebox{0.3\textwidth}{0.3\textwidth}
        {
          \begin{tikzpicture}[scale=0.07]
\def\colouredvertexscale{1.1}
\def\vertexscale{1.50}
\node [circle,black,draw,scale=\vertexscale] (1) at (-38.26834,-92.38795) {1};
\node [circle,black,draw,scale=\vertexscale] (2) at (-43.95169,-21.57490) {2};
\node [circle,black,draw,scale=\vertexscale] (3) at (-92.38795,38.26834) {3};
\node [circle,black,draw,scale=\vertexscale] (4) at (92.38795,-38.26834) {4};
\node [circle,black,draw,scale=\vertexscale] (5) at (34.24920,3.86246) {5};
\node [circle,black,draw,scale=\vertexscale] (6) at (55.88894,26.54927) {6};
\node [circle,black,draw,scale=\vertexscale] (7) at (17.64771,-24.93152) {7};
\node [circle,black,draw,scale=\vertexscale] (8) at (38.26834,92.38795) {8};
\node [circle,black,draw,scale=\vertexscale] (9) at (38.26834,-92.38795) {9};
\node [circle,black,draw,scale=0.9*\vertexscale] (10) at (22.82847,58.24753) {10};
\node [circle,black,draw,scale=0.9*\vertexscale] (11) at (-19.69432,44.80189) {11};
\node [circle,black,draw,scale=0.9*\vertexscale] (12) at (92.38795,38.26834) {12};
\node [circle,black,draw,scale=0.9*\vertexscale] (13) at (-92.38795,-38.26834) {13};
\node [circle,black,draw,scale=0.9*\vertexscale] (14) at (-38.26834,92.38795) {14};
\node [circle,black,draw,scale=0.9*\vertexscale] (15) at (-19.81030,-6.26457) {15};
\node [circle,black,draw,scale=0.9*\vertexscale] (16) at (39.72855,-47.08345) {16};
\node [circle,blue, thick, densely dotted,text=black,draw,scale=1.3*\colouredvertexscale] (17) at (-68.10281,-29.18666) {5};
\node [circle,blue, thick, densely dotted,text=black,draw,scale=1.3*\colouredvertexscale] (18) at (8.14516,22.86127) {13};
\draw [black] (1) to (2);
\draw [black] (1) to (9);
\draw [black] (1) to (13);
\draw [black] (2) to (3);
\draw [black] (2) to (15);
\draw [black] (3) to (13);
\draw [black] (3) to (14);
\draw [black] (4) to (9);
\draw [black] (4) to (16);
\draw [black] (4) to (12);
\draw [black] (5) to (6);
\draw [black] (5) to (7);
\draw [black] (5) to (18);
\draw [black] (6) to (10);
\draw [black] (6) to (12);
\draw [black] (7) to (16);
\draw [black] (7) to (15);
\draw [black] (8) to (14);
\draw [black] (8) to (12);
\draw [black] (8) to (10);
\draw [black] (9) to (16);
\draw [black] (10) to (11);
\draw [black] (11) to (15);
\draw [black] (11) to (14);
\draw [black] (13) to (17);
\end{tikzpicture}
          }
	\caption{An example of a cubic toroidal map drawn with disjoint nc-cycles. All faces are easily visible, including the face that generates a loop in the dual. It is the union of the two faces with the blue cycle -- that is
        $1\to 2\to 3\to 13\to 5\to 6\to 10\to 11\to 15\to 7\to 5\to 13\to 1$.}
	\label{fig:examplecubtorusT}
\end{figure}

So the computation of drawings with disjoint nc-cycles is more expensive than inside a fundamental polygon, but on the other hand it often gives very nice and
informative results -- e.g.\ for cubic toroidal maps.  Most of such maps allow very short nc-cycles: even restricting the counts to 3-connected graphs,
e.g. $64\%$ of the $398\ 871\ 372$ non-isomorphic 3-connected cubic toroidal maps on up to $24$ vertices have loops in the dual, so nc-cycles crossing only one
edge. Furthermore, even $99.8\%$ have double edges in the dual, so nc-cycles crossing two edges. In such drawings the faces can almost as easily be seen as in planar
maps. An example is given in Figure~\ref{fig:examplecubtorusT}.

\section{Drawings of abstract graphs}\label{sec:nonmap}

Planar\_draw draws only maps and is not designed for drawing abstract graphs. Nevertheless, in combination with programs embedding the graph to form a map, like e.g.\ {\em multi\_genus} described in \cite{genuscomp} or {\em
  multi\_allembed} described in \cite{allembed}, an embedding or even all embeddings can be computed and used as input for the drawing program. When interested
in abstract graphs, the exact map is not relevant, so the only criteria to choose for a certain map to be drawn are computing time and quality of
the drawing. Even when the emphasis is on computing time, one would probably not choose for a random map, but -- if possible -- go for a minimum genus embedding.
Computing the genus is NP-complete, so from the perspective of complexity theory, computing minimum genus embeddings is a time consuming task. While some NP-complete problems
can still be solved relatively fast for a large fraction of typical inputs, this is not the case for the genus problem. Nevertheless for many interesting graphs a minimum genus embedding can
still be computed in a reasonable amount of time.

\begin{figure}[p]
	\centering
        \resizebox{0.49\textwidth}{0.49\textwidth}
        {
          \input{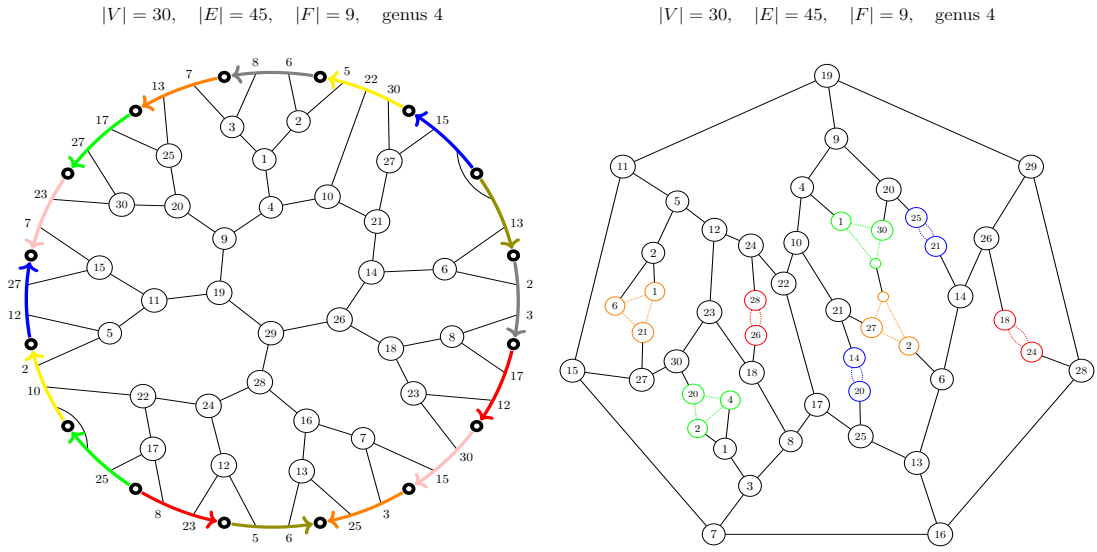}
        }
               \resizebox{0.49\textwidth}{0.49\textwidth}
        {
          \begin{tikzpicture}[scale=0.07]
\def\colouredvertexscale{0.86}
\def\vertexscale{0.99}
\node [circle,black,draw,scale=\vertexscale] (1) at (-39.06910,-54.59900) {1};
\node [circle,black,draw,scale=\vertexscale] (2) at (-66.53826,26.60238) {2};
\node [circle,black,draw,scale=\vertexscale] (3) at (-29.48147,-69.99476) {3};
\node [circle,black,draw,scale=\vertexscale] (4) at (-9.49711,53.77899) {4};
\node [circle,black,draw,scale=\vertexscale] (5) at (-57.29782,47.89383) {5};
\node [circle,black,draw,scale=\vertexscale] (6) at (44.05158,-25.81414) {6};
\node [circle,black,draw,scale=\vertexscale] (7) at (-43.38837,-90.09689) {7};
\node [circle,black,draw,scale=\vertexscale] (8) at (-13.85535,-51.49238) {8};
\node [circle,black,draw,scale=\vertexscale] (9) at (3.58538,73.96045) {9};
\node [circle,black,draw,scale=0.9*\vertexscale] (10) at (-11.74958,30.75586) {10};
\node [circle,black,draw,scale=0.9*\vertexscale] (11) at (-78.18315,62.34898) {11};
\node [circle,black,draw,scale=0.9*\vertexscale] (12) at (-43.09141,36.15939) {12};
\node [circle,black,draw,scale=0.9*\vertexscale] (13) at (34.39486,-60.51783) {13};
\node [circle,black,draw,scale=0.9*\vertexscale] (14) at (51.13796,8.56085) {14};
\node [circle,black,draw,scale=0.9*\vertexscale] (15) at (-97.49279,-22.25209) {15};
\node [circle,black,draw,scale=0.9*\vertexscale] (16) at (43.38837,-90.09689) {16};
\node [circle,black,draw,scale=0.9*\vertexscale] (17) at (-3.80252,-36.39816) {17};
\node [circle,black,draw,scale=0.9*\vertexscale] (18) at (-28.78031,-23.11640) {18};
\node [circle,black,draw,scale=0.9*\vertexscale] (19) at (-0.00000,100.00000) {19};
\node [circle,black,draw,scale=0.9*\vertexscale] (20) at (23.65879,52.86243) {20};
\node [circle,black,draw,scale=0.9*\vertexscale] (21) at (4.32000,2.87212) {21};
\node [circle,black,draw,scale=0.9*\vertexscale] (22) at (-16.60434,13.87533) {22};
\node [circle,black,draw,scale=0.9*\vertexscale] (23) at (-45.00682,2.01470) {23};
\node [circle,black,draw,scale=0.9*\vertexscale] (24) at (-29.14113,29.65189) {24};
\node [circle,black,draw,scale=0.9*\vertexscale] (25) at (12.92391,-49.47911) {25};
\node [circle,black,draw,scale=0.9*\vertexscale] (26) at (61.09964,32.61670) {26};
\node [circle,black,draw,scale=0.9*\vertexscale] (27) at (-70.96377,-26.03270) {27};
\node [circle,black,draw,scale=0.9*\vertexscale] (28) at (97.49279,-22.25209) {28};
\node [circle,black,draw,scale=0.9*\vertexscale] (29) at (78.18315,62.34898) {29};
\node [circle,black,draw,scale=0.9*\vertexscale] (30) at (-57.64082,-18.37501) {30};
\node [circle,blue,text=black,draw,scale=0.9*\colouredvertexscale] (31) at (10.50047,-16.86848) {14};
\node [circle,orange,text=black,draw,scale=\colouredvertexscale] (32) at (-65.93013,10.54710) {1};
\node [circle,blue,text=black,draw,scale=0.9*\colouredvertexscale] (33) at (12.49398,-30.81379) {20};
\node [circle,red,text=black,draw,scale=0.9*\colouredvertexscale] (34) at (68.12803,-1.25794) {18};
\node [circle,orange,text=black,draw,scale=0.9*\colouredvertexscale] (35) at (-70.49448,-6.40654) {21};
\node [circle,red,text=black,draw,scale=0.9*\colouredvertexscale] (36) at (78.43777,-14.53665) {24};
\node [circle,draw, green,scale=\colouredvertexscale] (37) at (18.57274,22.19679) {};
\node [circle,orange,text=black,draw,scale=\colouredvertexscale] (38) at (-81.21031,4.57381) {6};
\node [circle,green,text=black,draw,scale=\colouredvertexscale] (39) at (5.41629,39.73373) {1};
\node [circle,green,text=black,draw,scale=0.9*\colouredvertexscale] (40) at (21.13740,35.96305) {30};
\node [circle,orange,text=black,draw,scale=\colouredvertexscale] (41) at (31.39558,-12.01626) {2};
\node [circle,green,text=black,draw,scale=0.9*\colouredvertexscale] (42) at (-51.20822,-31.94205) {20};
\node [circle,orange,text=black,draw,scale=0.9*\colouredvertexscale] (43) at (17.10653,-4.39301) {27};
\node [circle,draw, orange,scale=\colouredvertexscale] (44) at (21.56048,8.38655) {};
\node [circle,blue,text=black,draw,scale=0.9*\colouredvertexscale] (45) at (41.70216,29.27578) {21};
\node [circle,blue,text=black,draw,scale=0.9*\colouredvertexscale] (46) at (34.22761,41.25566) {25};
\node [circle,green,text=black,draw,scale=\colouredvertexscale] (47) at (-37.03597,-34.45503) {4};
\node [circle,red,text=black,draw,scale=0.9*\colouredvertexscale] (48) at (-27.23510,-7.60096) {26};
\node [circle,red,text=black,draw,scale=0.9*\colouredvertexscale] (49) at (-27.44778,6.89494) {28};
\node [circle,green,text=black,draw,scale=\colouredvertexscale] (50) at (-49.55955,-46.20321) {2};
\draw [black] (1) to (3);
\draw [black] (1) to (50);
\draw [black] (1) to (47);
\draw [black] (2) to (5);
\draw [black] (2) to (32);
\draw [black] (2) to (38);
\draw [black] (3) to (7);
\draw [black] (3) to (8);
\draw [black] (4) to (9);
\draw [black] (4) to (39);
\draw [black] (4) to (10);
\draw [black] (5) to (12);
\draw [black] (5) to (11);
\draw [black] (6) to (14);
\draw [black] (6) to (13);
\draw [black] (6) to (41);
\draw [black] (7) to (16);
\draw [black] (7) to (15);
\draw [black] (8) to (18);
\draw [black] (8) to (17);
\draw [black] (9) to (19);
\draw [black] (9) to (20);
\draw [black] (10) to (21);
\draw [black] (10) to (22);
\draw [black] (11) to (19);
\draw [black] (11) to (15);
\draw [black] (12) to (24);
\draw [black] (12) to (23);
\draw [black] (13) to (16);
\draw [black] (13) to (25);
\draw [black] (14) to (26);
\draw [black] (14) to (45);
\draw [black] (15) to (27);
\draw [black] (16) to (28);
\draw [black] (17) to (22);
\draw [black] (17) to (25);
\draw [black] (18) to (48);
\draw [black] (18) to (23);
\draw [black] (19) to (29);
\draw [black] (20) to (40);
\draw [black] (20) to (46);
\draw [black] (21) to (43);
\draw [black] (21) to (31);
\draw [black] (22) to (24);
\draw [black] (23) to (30);
\draw [black] (24) to (49);
\draw [black] (25) to (33);
\draw [black] (26) to (34);
\draw [black] (26) to (29);
\draw [black] (27) to (30);
\draw [black] (27) to (35);
\draw [black] (28) to (36);
\draw [black] (28) to (29);
\draw [black] (30) to (42);
\draw [blue, bend right=20, thick, densely dotted] (31) to (33);
\draw [blue, bend left=20, thick, densely dotted] (31) to (33);
\draw [orange, thick, densely dotted] (32) to (35);
\draw [orange, thick, densely dotted] (32) to (38);
\draw [red, bend right=20, thick, densely dotted] (34) to (36);
\draw [red, bend left=20, thick, densely dotted] (34) to (36);
\draw [orange, thick, densely dotted] (35) to (38);
\draw [black] (37) to (44);
\draw [green, thick, densely dotted] (37) to (39);
\draw [green, thick, densely dotted] (37) to (40);
\draw [green, thick, densely dotted] (39) to (40);
\draw [orange, thick, densely dotted] (41) to (43);
\draw [orange, thick, densely dotted] (41) to (44);
\draw [green, thick, densely dotted] (42) to (47);
\draw [green, thick, densely dotted] (42) to (50);
\draw [orange, thick, densely dotted] (43) to (44);
\draw [blue, bend right=20, thick, densely dotted] (45) to (46);
\draw [blue, bend left=20, thick, densely dotted] (45) to (46);
\draw [green, thick, densely dotted] (47) to (50);
\draw [red, bend right=20, thick, densely dotted] (48) to (49);
\draw [red, bend left=20, thick, densely dotted] (48) to (49);
\draw[color=black] (0.0,125.00000) node [font={\Large}] {$|V|=30, \quad   |E|=45,    \quad  |F|=9,    \quad  \mbox{genus } 4$};
\end{tikzpicture}
          }
	\caption{Minimum genus drawings of an abstract cubic graph on 30 vertices.}
	\label{fig:manyT30}
\end{figure}

For not too large graphs with a not too large genus, one could generate one embedding and draw that. If computing time is less important (e.g. producing a drawing for an article),
and the restrictions on size and genus are even stronger, one can generate
all minimum genus embeddings and choose one drawing among all these embeddings. Both is supported by planar\_draw, but one must take into account that some
graphs have a huge number of minimum genus embeddings, which makes it in combination with the difficulty of computing a single minimum genus embedding especially hard. We give two examples:

Figure~\ref{fig:manyT30} shows two drawings of an abstract cubic graph on 30 vertices.  Computing just one minimum genus map of this graph and drawing it took $0.1$ seconds
for drawings inside a fundamental polygon and $0.2$ seconds for a drawing with disjoint nc-cycles.  The drawings in Figure~\ref{fig:manyT30} were chosen from
drawings of all non-isomorphic minimum genus maps.  For these drawings, the $3\ 336$ non-isomorphic maps of genus $4$ were computed by multi\_allembed, cut open and drawn in
several ways until finally the displayed ones were chosen.  For the drawings inside a fundamental polygon in total $4\ 546$ drawings were produced and the
computation including the generation of all minimum genus maps and computing all drawings took $3$ seconds.  For the drawings with disjoint nc-cycles in total
$1\ 532\ 408$ drawings were produced and the computation including the generation of all minimum genus maps and computing all drawings took $20$ minutes.  Note
that the maps chosen in these two cases are not isomorphic.

\begin{figure}[p]
	\centering
        \resizebox{0.49\textwidth}{0.49\textwidth}
        {
          \input{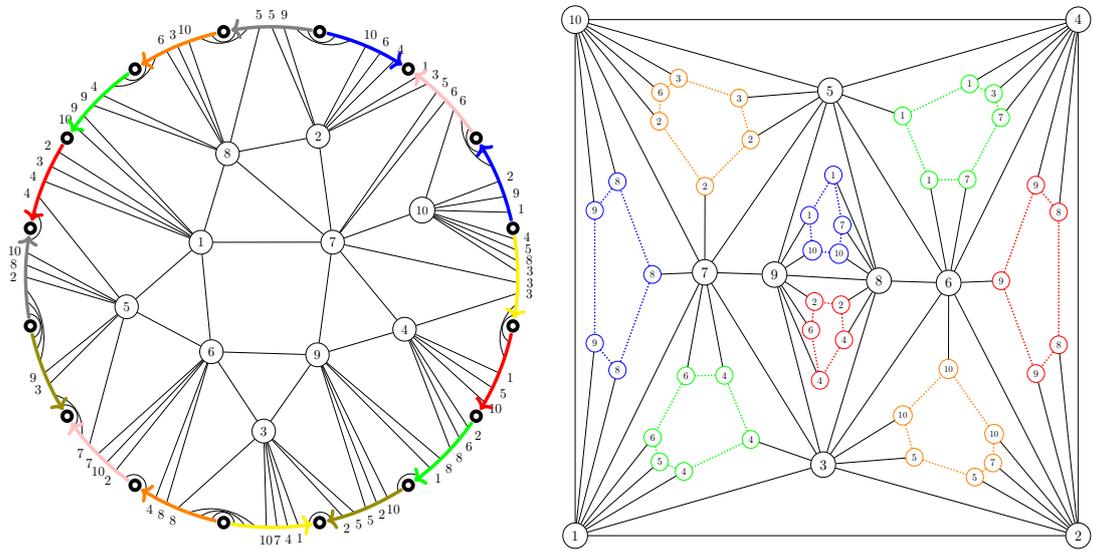}
        }
               \resizebox{0.49\textwidth}{0.49\textwidth}
        {
          \begin{tikzpicture}[scale=0.07]
\def\colouredvertexscale{0.55}
\def\vertexscale{0.79}
\node [circle,black,draw,scale=\vertexscale] (1) at (-70.71068,-70.71068) {1};
\node [circle,black,draw,scale=\vertexscale] (2) at (70.71068,-70.71068) {2};
\node [circle,black,draw,scale=\vertexscale] (3) at (-0.98681,-51.24489) {3};
\node [circle,black,draw,scale=\vertexscale] (4) at (70.71068,70.71068) {4};
\node [circle,black,draw,scale=\vertexscale] (5) at (0.98681,51.24489) {5};
\node [circle,black,draw,scale=\vertexscale] (6) at (34.28765,-1.44612) {6};
\node [circle,black,draw,scale=\vertexscale] (7) at (-34.28765,1.44612) {7};
\node [circle,black,draw,scale=\vertexscale] (8) at (14.67658,-0.83643) {8};
\node [circle,black,draw,scale=\vertexscale] (9) at (-14.67658,0.83643) {9};
\node [circle,black,draw,scale=0.9*\vertexscale] (10) at (-70.71068,70.71068) {10};
\node [circle,blue,text=black,draw,scale=\colouredvertexscale] (11) at (1.88195,28.13405) {1};
\node [circle,orange,text=black,draw,scale=\colouredvertexscale] (12) at (24.64116,-49.22806) {5};
\node [circle,blue,text=black,draw,scale=\colouredvertexscale] (13) at (4.41196,14.39188) {7};
\node [circle,blue,text=black,draw,scale=0.9*\colouredvertexscale] (14) at (3.42632,6.55305) {10};
\node [circle,blue,text=black,draw,scale=0.9*\colouredvertexscale] (15) at (-4.02259,7.44575) {10};
\node [circle,blue,text=black,draw,scale=\colouredvertexscale] (16) at (-4.86333,17.06362) {1};
\node [circle,red,text=black,draw,scale=\colouredvertexscale] (17) at (4.02259,-7.44575) {2};
\node [circle,orange,text=black,draw,scale=0.9*\colouredvertexscale] (18) at (21.39243,-37.76062) {10};
\node [circle,red,text=black,draw,scale=\colouredvertexscale] (19) at (4.86333,-17.06362) {4};
\node [circle,red,text=black,draw,scale=\colouredvertexscale] (20) at (-1.88195,-28.13405) {4};
\node [circle,red,text=black,draw,scale=\colouredvertexscale] (21) at (-4.41196,-14.39188) {6};
\node [circle,red,text=black,draw,scale=\colouredvertexscale] (22) at (-3.42632,-6.55305) {2};
\node [circle,green,text=black,draw,scale=\colouredvertexscale] (23) at (40.18371,53.10874) {1};
\node [circle,orange,text=black,draw,scale=0.9*\colouredvertexscale] (24) at (34.21926,-25.03102) {10};
\node [circle,green,text=black,draw,scale=\colouredvertexscale] (25) at (46.96746,50.47325) {3};
\node [circle,green,text=black,draw,scale=\colouredvertexscale] (26) at (48.94570,43.85940) {7};
\node [circle,green,text=black,draw,scale=\colouredvertexscale] (27) at (39.56365,26.90798) {7};
\node [circle,green,text=black,draw,scale=\colouredvertexscale] (28) at (28.78444,26.75258) {1};
\node [circle,green,text=black,draw,scale=\colouredvertexscale] (29) at (21.33318,44.33849) {1};
\node [circle,orange,text=black,draw,scale=\colouredvertexscale] (30) at (-21.39243,37.76062) {2};
\node [circle,orange,text=black,draw,scale=0.9*\colouredvertexscale] (31) at (47.07329,-42.84918) {10};
\node [circle,orange,text=black,draw,scale=\colouredvertexscale] (32) at (-34.21926,25.03102) {2};
\node [circle,orange,text=black,draw,scale=\colouredvertexscale] (33) at (-47.07329,42.84918) {2};
\node [circle,orange,text=black,draw,scale=\colouredvertexscale] (34) at (-46.74335,50.78748) {6};
\node [circle,orange,text=black,draw,scale=\colouredvertexscale] (35) at (-41.65161,54.65829) {3};
\node [circle,orange,text=black,draw,scale=\colouredvertexscale] (36) at (-24.64116,49.22806) {3};
\node [circle,blue,text=black,draw,scale=\colouredvertexscale] (37) at (-58.81852,-25.35208) {8};
\node [circle,blue,text=black,draw,scale=\colouredvertexscale] (38) at (-49.01413,0.86197) {8};
\node [circle,blue,text=black,draw,scale=\colouredvertexscale] (39) at (-58.81839,26.29779) {8};
\node [circle,blue,text=black,draw,scale=\colouredvertexscale] (40) at (-65.21066,18.41242) {9};
\node [circle,blue,text=black,draw,scale=\colouredvertexscale] (41) at (-65.21066,-17.97395) {9};
\node [circle,orange,text=black,draw,scale=\colouredvertexscale] (42) at (46.74335,-50.78748) {7};
\node [circle,red,text=black,draw,scale=\colouredvertexscale] (43) at (65.21066,-18.41242) {8};
\node [circle,red,text=black,draw,scale=\colouredvertexscale] (44) at (65.21066,17.97395) {8};
\node [circle,red,text=black,draw,scale=\colouredvertexscale] (45) at (58.81852,25.35208) {9};
\node [circle,red,text=black,draw,scale=\colouredvertexscale] (46) at (49.01413,-0.86197) {9};
\node [circle,red,text=black,draw,scale=\colouredvertexscale] (47) at (58.81839,-26.29779) {9};
\node [circle,orange,text=black,draw,scale=\colouredvertexscale] (48) at (41.65161,-54.65829) {5};
\node [circle,green,text=black,draw,scale=\colouredvertexscale] (49) at (-40.18371,-53.10874) {4};
\node [circle,green,text=black,draw,scale=\colouredvertexscale] (50) at (-21.33318,-44.33849) {4};
\node [circle,green,text=black,draw,scale=\colouredvertexscale] (51) at (-28.78444,-26.75258) {4};
\node [circle,green,text=black,draw,scale=\colouredvertexscale] (52) at (-39.56365,-26.90798) {6};
\node [circle,green,text=black,draw,scale=\colouredvertexscale] (53) at (-48.94570,-43.85940) {6};
\node [circle,green,text=black,draw,scale=\colouredvertexscale] (54) at (-46.96746,-50.47325) {5};
\draw [black] (1) to (2);
\draw [black] (1) to (10);
\draw [black] (1) to (41);
\draw [black] (1) to (37);
\draw [black] (1) to (7);
\draw [black] (1) to (53);
\draw [black] (1) to (54);
\draw [black] (1) to (49);
\draw [black] (1) to (3);
\draw [black] (2) to (3);
\draw [black] (2) to (48);
\draw [black] (2) to (42);
\draw [black] (2) to (31);
\draw [black] (2) to (6);
\draw [black] (2) to (47);
\draw [black] (2) to (43);
\draw [black] (2) to (4);
\draw [black] (3) to (50);
\draw [black] (3) to (7);
\draw [black] (3) to (9);
\draw [black] (3) to (8);
\draw [black] (3) to (6);
\draw [black] (3) to (18);
\draw [black] (3) to (12);
\draw [black] (4) to (23);
\draw [black] (4) to (5);
\draw [black] (4) to (10);
\draw [black] (4) to (44);
\draw [black] (4) to (45);
\draw [black] (4) to (6);
\draw [black] (4) to (26);
\draw [black] (4) to (25);
\draw [black] (5) to (30);
\draw [black] (5) to (36);
\draw [black] (5) to (10);
\draw [black] (5) to (29);
\draw [black] (5) to (6);
\draw [black] (5) to (8);
\draw [black] (5) to (9);
\draw [black] (5) to (7);
\draw [black] (6) to (28);
\draw [black] (6) to (27);
\draw [black] (6) to (46);
\draw [black] (6) to (24);
\draw [black] (6) to (8);
\draw [black] (7) to (52);
\draw [black] (7) to (38);
\draw [black] (7) to (10);
\draw [black] (7) to (32);
\draw [black] (7) to (9);
\draw [black] (7) to (51);
\draw [black] (8) to (13);
\draw [black] (8) to (11);
\draw [black] (8) to (19);
\draw [black] (8) to (17);
\draw [black] (8) to (9);
\draw [black] (8) to (14);
\draw [black] (9) to (22);
\draw [black] (9) to (21);
\draw [black] (9) to (20);
\draw [black] (9) to (16);
\draw [black] (9) to (15);
\draw [black] (10) to (40);
\draw [black] (10) to (35);
\draw [black] (10) to (34);
\draw [black] (10) to (33);
\draw [black] (10) to (39);
\draw [blue, thick, densely dotted] (11) to (13);
\draw [blue, thick, densely dotted] (11) to (16);
\draw [orange, thick, densely dotted] (12) to (48);
\draw [orange, thick, densely dotted] (12) to (18);
\draw [blue, thick, densely dotted] (13) to (14);
\draw [blue, thick, densely dotted] (14) to (15);
\draw [blue, thick, densely dotted] (15) to (16);
\draw [red, thick, densely dotted] (17) to (19);
\draw [red, thick, densely dotted] (17) to (22);
\draw [orange, thick, densely dotted] (18) to (24);
\draw [red, thick, densely dotted] (19) to (20);
\draw [red, thick, densely dotted] (20) to (21);
\draw [red, thick, densely dotted] (21) to (22);
\draw [green, thick, densely dotted] (23) to (25);
\draw [green, thick, densely dotted] (23) to (29);
\draw [orange, thick, densely dotted] (24) to (31);
\draw [green, thick, densely dotted] (25) to (26);
\draw [green, thick, densely dotted] (26) to (27);
\draw [green, thick, densely dotted] (27) to (28);
\draw [green, thick, densely dotted] (28) to (29);
\draw [orange, thick, densely dotted] (30) to (32);
\draw [orange, thick, densely dotted] (30) to (36);
\draw [orange, thick, densely dotted] (31) to (42);
\draw [orange, thick, densely dotted] (32) to (33);
\draw [orange, thick, densely dotted] (33) to (34);
\draw [orange, thick, densely dotted] (34) to (35);
\draw [orange, thick, densely dotted] (35) to (36);
\draw [blue, thick, densely dotted] (37) to (38);
\draw [blue, thick, densely dotted] (37) to (41);
\draw [blue, thick, densely dotted] (38) to (39);
\draw [blue, thick, densely dotted] (39) to (40);
\draw [blue, thick, densely dotted] (40) to (41);
\draw [orange, thick, densely dotted] (42) to (48);
\draw [red, thick, densely dotted] (43) to (44);
\draw [red, thick, densely dotted] (43) to (47);
\draw [red, thick, densely dotted] (44) to (45);
\draw [red, thick, densely dotted] (45) to (46);
\draw [red, thick, densely dotted] (46) to (47);
\draw [green, thick, densely dotted] (49) to (50);
\draw [green, thick, densely dotted] (49) to (54);
\draw [green, thick, densely dotted] (50) to (51);
\draw [green, thick, densely dotted] (51) to (52);
\draw [green, thick, densely dotted] (52) to (53);
\draw [green, thick, densely dotted] (53) to (54);
\end{tikzpicture}
          }
	\caption{An expensive case when choosing from drawings of all non-isomorphic minimum genus embeddings: minimum genus drawings of $K_{10}$.}   
	\label{fig:manyTk10}
\end{figure}

Figure~\ref{fig:manyTk10} shows two drawings of $K_{10}$.  Computing just one minimum genus map and drawing it takes altogether $0.07$ seconds for a drawing in
a fundamental polygon and $1.8$ seconds for drawing $K_{10}$ with disjoint nc-cycles.  The drawings in Figure~\ref{fig:manyTk10} were chosen from drawings of
all non-isomorphic maps.  For the drawings, all $1\ 083$ non-isomorphic maps of genus 4 were computed. This took $38.5$ hours. Having the maps available, to find
{\em the best} drawing inside a fundamental polygon, $1\ 084$ drawings were produced (so one more than maps) and the one on the left was chosen. This took $1.3$
seconds. For {\em the best} drawing with disjoint nc-cycles, in total $1\ 074\ 721$ drawings were produced, which took $27$ minutes.

\section{Conclusion}

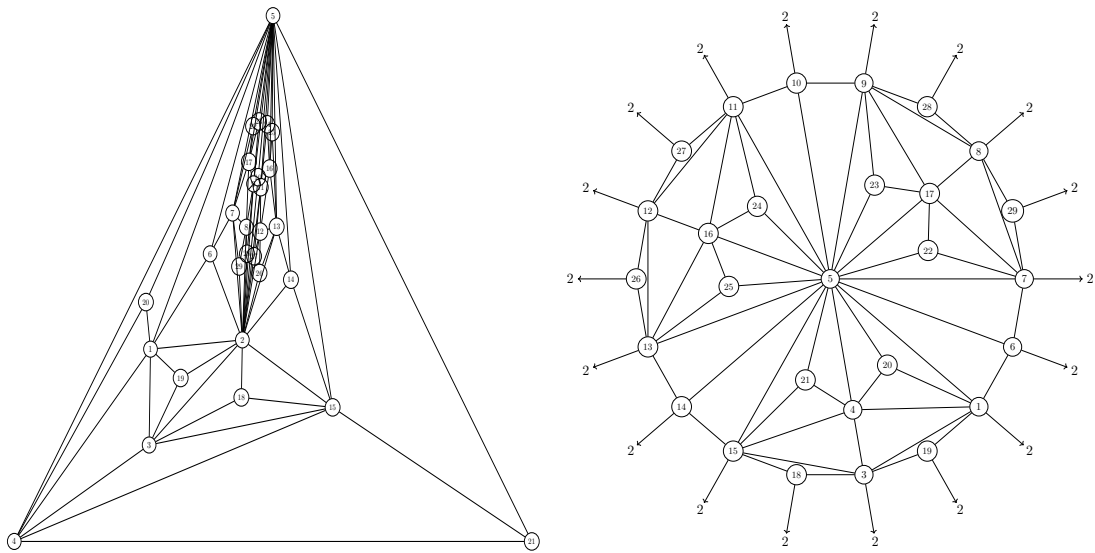
\begin{figure}[p]
	\centering
        \resizebox{0.49\textwidth}{0.49\textwidth}
        {
          \begin{tikzpicture}[scale=0.07]
\def\vertexscale{0.52}
\node [circle,black,draw,scale=\vertexscale] (1) at (-41.00358,4.85797) {1};
\node [circle,black,draw,scale=\vertexscale] (2) at (-10.25932,7.49393) {2};
\node [circle,black,draw,scale=\vertexscale] (3) at (-41.46572,-22.48869) {3};
\node [circle,black,draw,scale=\vertexscale] (4) at (-86.60254,-50.00000) {4};
\node [circle,black,draw,scale=\vertexscale] (5) at (-0.00000,100.00000) {5};
\node [circle,black,draw,scale=\vertexscale] (6) at (-21.05362,32.04736) {6};
\node [circle,black,draw,scale=\vertexscale] (7) at (-13.54122,43.69149) {7};
\node [circle,black,draw,scale=\vertexscale] (8) at (-8.93973,39.64382) {8};
\node [circle,black,draw,scale=\vertexscale] (9) at (-6.53468,52.00601) {9};
\node [circle,black,draw,scale=0.9*\vertexscale] (10) at (-4.97928,53.91882) {10};
\node [circle,black,draw,scale=0.9*\vertexscale] (11) at (-4.10873,51.06753) {11};
\node [circle,black,draw,scale=0.9*\vertexscale] (12) at (-4.26591,38.39460) {12};
\node [circle,black,draw,scale=0.9*\vertexscale] (13) at (1.21993,39.78918) {13};
\node [circle,black,draw,scale=0.9*\vertexscale] (14) at (6.00368,24.66501) {14};
\node [circle,black,draw,scale=0.9*\vertexscale] (15) at (19.94587,-11.75086) {15};
\node [circle,black,draw,scale=0.9*\vertexscale] (16) at (-1.12836,56.39761) {16};
\node [circle,black,draw,scale=0.9*\vertexscale] (17) at (-8.09321,58.29561) {17};
\node [circle,black,draw,scale=0.9*\vertexscale] (18) at (-10.59334,-8.91530) {18};
\node [circle,black,draw,scale=0.9*\vertexscale] (19) at (-30.90899,-3.37828) {19};
\node [circle,black,draw,scale=0.9*\vertexscale] (20) at (-42.53532,18.28620) {20};
\node [circle,black,draw,scale=0.9*\vertexscale] (21) at (86.60254,-50.00000) {21};
\node [circle,black,draw,scale=0.9*\vertexscale] (22) at (-6.67904,68.47927) {22};
\node [circle,black,draw,scale=0.9*\vertexscale] (23) at (-4.71039,69.82563) {23};
\node [circle,black,draw,scale=0.9*\vertexscale] (24) at (-1.95930,69.01616) {24};
\node [circle,black,draw,scale=0.9*\vertexscale] (25) at (-0.30201,66.74805) {25};
\node [circle,black,draw,scale=0.9*\vertexscale] (26) at (-4.52910,26.57710) {26};
\node [circle,black,draw,scale=0.9*\vertexscale] (27) at (-6.35294,31.37216) {27};
\node [circle,black,draw,scale=0.9*\vertexscale] (28) at (-8.71871,32.08315) {28};
\node [circle,black,draw,scale=0.9*\vertexscale] (29) at (-11.35682,28.44328) {29};
\draw [black] (1) to (6);
\draw [black] (1) to (2);
\draw [black] (1) to (19);
\draw [black] (1) to (3);
\draw [black] (1) to (4);
\draw [black] (1) to (20);
\draw [black] (1) to (5);
\draw [black] (2) to (15);
\draw [black] (2) to (18);
\draw [black] (2) to (3);
\draw [black] (2) to (19);
\draw [black] (2) to (6);
\draw [black] (2) to (7);
\draw [black] (2) to (29);
\draw [black] (2) to (8);
\draw [black] (2) to (28);
\draw [black] (2) to (9);
\draw [black] (2) to (10);
\draw [black] (2) to (11);
\draw [black] (2) to (27);
\draw [black] (2) to (12);
\draw [black] (2) to (26);
\draw [black] (2) to (13);
\draw [black] (2) to (14);
\draw [black] (3) to (18);
\draw [black] (3) to (15);
\draw [black] (3) to (4);
\draw [black] (3) to (19);
\draw [black] (4) to (20);
\draw [black] (4) to (15);
\draw [black] (4) to (21);
\draw [black] (4) to (5);
\draw [black] (5) to (20);
\draw [black] (5) to (21);
\draw [black] (5) to (15);
\draw [black] (5) to (14);
\draw [black] (5) to (13);
\draw [black] (5) to (25);
\draw [black] (5) to (16);
\draw [black] (5) to (24);
\draw [black] (5) to (11);
\draw [black] (5) to (10);
\draw [black] (5) to (9);
\draw [black] (5) to (23);
\draw [black] (5) to (17);
\draw [black] (5) to (22);
\draw [black] (5) to (7);
\draw [black] (5) to (6);
\draw [black] (6) to (7);
\draw [black] (7) to (22);
\draw [black] (7) to (17);
\draw [black] (7) to (8);
\draw [black] (7) to (29);
\draw [black] (8) to (29);
\draw [black] (8) to (17);
\draw [black] (8) to (9);
\draw [black] (8) to (28);
\draw [black] (9) to (28);
\draw [black] (9) to (17);
\draw [black] (9) to (23);
\draw [black] (9) to (10);
\draw [black] (10) to (11);
\draw [black] (11) to (24);
\draw [black] (11) to (16);
\draw [black] (11) to (12);
\draw [black] (11) to (27);
\draw [black] (12) to (27);
\draw [black] (12) to (16);
\draw [black] (12) to (13);
\draw [black] (12) to (26);
\draw [black] (13) to (26);
\draw [black] (13) to (16);
\draw [black] (13) to (25);
\draw [black] (13) to (14);
\draw [black] (14) to (15);
\draw [black] (15) to (18);
\draw [black] (15) to (21);
\draw [black] (16) to (24);
\draw [black] (16) to (25);
\draw [black] (17) to (22);
\draw [black] (17) to (23);
\end{tikzpicture}
        }
              \resizebox{0.49\textwidth}{0.49\textwidth}
        {
          \begin{tikzpicture}[scale=0.06]
\def\vertexscale{0.91}
\def\labelscale{1.21}
\node [circle,black,draw,scale=\vertexscale] (1) at (76.60444,-64.27876) {1};
\node [circle,black,draw,scale=\vertexscale] (2) at (93.96926,34.20201) {29};
\node [circle,black,draw,scale=\vertexscale] (3) at (17.36482,-98.48078) {3};
\node [circle,black,draw,scale=\vertexscale] (4) at (11.61083,-65.84828) {4};
\node [circle,black,draw,scale=\vertexscale] (5) at (0.00000,-0.00000) {5};
\node [circle,black,draw,scale=\vertexscale] (6) at (93.96926,-34.20201) {6};
\node [circle,black,draw,scale=\vertexscale] (7) at (100.00000,0.00000) {7};
\node [circle,black,draw,scale=\vertexscale] (8) at (76.60444,64.27876) {8};
\node [circle,black,draw,scale=\vertexscale] (9) at (17.36482,98.48078) {9};
\node [circle,black,draw,scale=0.9*\vertexscale] (10) at (-17.36482,98.48078) {10};
\node [circle,black,draw,scale=0.9*\vertexscale] (11) at (-50.00000,86.60254) {11};
\node [circle,black,draw,scale=0.9*\vertexscale] (12) at (-93.96926,34.20201) {12};
\node [circle,black,draw,scale=0.9*\vertexscale] (13) at (-93.96926,-34.20201) {13};
\node [circle,black,draw,scale=0.9*\vertexscale] (14) at (-76.60444,-64.27876) {14};
\node [circle,black,draw,scale=0.9*\vertexscale] (15) at (-50.00000,-86.60254) {15};
\node [circle,black,draw,scale=0.9*\vertexscale] (16) at (-62.83170,22.86887) {16};
\node [circle,black,draw,scale=0.9*\vertexscale] (17) at (51.22087,42.97942) {17};
\node [circle,black,draw,scale=0.9*\vertexscale] (18) at (-17.36482,-98.48078) {18};
\node [circle,black,draw,scale=0.9*\vertexscale] (19) at (50.00000,-86.60254) {19};
\node [circle,black,draw,scale=0.9*\vertexscale] (20) at (29.41425,-43.36384) {20};
\node [circle,black,draw,scale=0.9*\vertexscale] (21) at (-12.80905,-50.80895) {21};
\node [circle,black,draw,scale=0.9*\vertexscale] (22) at (50.40637,14.31151) {22};
\node [circle,black,draw,scale=0.9*\vertexscale] (23) at (22.84706,47.15541) {23};
\node [circle,black,draw,scale=0.9*\vertexscale] (24) at (-37.59732,36.49744) {24};
\node [circle,black,draw,scale=0.9*\vertexscale] (25) at (-52.26132,-3.79157) {25};
\node [circle,black,draw,scale=0.9*\vertexscale] (26) at (-100.00000,-0.00000) {26};
\node [circle,black,draw,scale=0.9*\vertexscale] (27) at (-76.60444,64.27876) {27};
\node [circle,black,draw,scale=0.9*\vertexscale] (28) at (50.00000,86.60254) {28};
\draw [black] (1) to (19);
\draw [black] (1) to (3);
\draw [black] (1) to (4);
\draw [black] (1) to (20);
\draw [black] (1) to (5);
\draw [black] (1) to (6);
\draw [black] (2) to (7);
\draw [black] (2) to (8);
\draw [black] (3) to (18);
\draw [black] (3) to (15);
\draw [black] (3) to (4);
\draw [black] (3) to (19);
\draw [black] (4) to (15);
\draw [black] (4) to (21);
\draw [black] (4) to (5);
\draw [black] (4) to (20);
\draw [black] (5) to (21);
\draw [black] (5) to (15);
\draw [black] (5) to (14);
\draw [black] (5) to (13);
\draw [black] (5) to (25);
\draw [black] (5) to (16);
\draw [black] (5) to (24);
\draw [black] (5) to (11);
\draw [black] (5) to (10);
\draw [black] (5) to (9);
\draw [black] (5) to (23);
\draw [black] (5) to (17);
\draw [black] (5) to (22);
\draw [black] (5) to (7);
\draw [black] (5) to (6);
\draw [black] (5) to (20);
\draw [black] (6) to (7);
\draw [black] (7) to (22);
\draw [black] (7) to (17);
\draw [black] (7) to (8);
\draw [black] (8) to (17);
\draw [black] (8) to (9);
\draw [black] (8) to (28);
\draw [black] (9) to (17);
\draw [black] (9) to (23);
\draw [black] (9) to (10);
\draw [black] (9) to (28);
\draw [black] (10) to (11);
\draw [black] (11) to (24);
\draw [black] (11) to (16);
\draw [black] (11) to (12);
\draw [black] (11) to (27);
\draw [black] (12) to (16);
\draw [black] (12) to (13);
\draw [black] (12) to (26);
\draw [black] (12) to (27);
\draw [black] (13) to (16);
\draw [black] (13) to (25);
\draw [black] (13) to (14);
\draw [black] (13) to (26);
\draw [black] (14) to (15);
\draw [black] (15) to (18);
\draw [black] (15) to (21);
\draw [black] (16) to (25);
\draw [black] (16) to (24);
\draw [black] (17) to (22);
\draw [black] (17) to (23);
\draw [black,->] (18) to (-22.57426,-128.02501);
\node[draw=none,fill=none,scale=\labelscale] () at (-23.26886,-131.96424) {2};
\draw [black,->] (15) to (-65.00000,-112.58330);
\node[draw=none,fill=none,scale=\labelscale] () at (-67.00000,-116.04740) {2};
\draw [black,->] (14) to (-99.58578,-83.56239);
\node[draw=none,fill=none,scale=\labelscale] () at (-102.64996,-86.13354) {2};
\draw [black,->] (13) to (-122.16004,-44.46262);
\node[draw=none,fill=none,scale=\labelscale] () at (-125.91881,-45.83070) {2};
\draw [black,->] (26) to (-130.00000,-0.00000);
\node[draw=none,fill=none,scale=\labelscale] () at (-134.00000,-0.00000) {2};
\draw [black,->] (12) to (-122.16004,44.46262);
\node[draw=none,fill=none,scale=\labelscale] () at (-125.91881,45.83070) {2};
\draw [black,->] (27) to (-99.58578,83.56239);
\node[draw=none,fill=none,scale=\labelscale] () at (-102.64996,86.13354) {2};
\draw [black,->] (11) to (-65.00000,112.58330);
\node[draw=none,fill=none,scale=\labelscale] () at (-67.00000,116.04740) {2};
\draw [black,->] (10) to (-22.57426,128.02501);
\node[draw=none,fill=none,scale=\labelscale] () at (-23.26886,131.96424) {2};
\draw [black,->] (9) to (22.57426,128.02501);
\node[draw=none,fill=none,scale=\labelscale] () at (23.26886,131.96424) {2};
\draw [black,->] (28) to (65.00000,112.58330);
\node[draw=none,fill=none,scale=\labelscale] () at (67.00000,116.04740) {2};
\draw [black,->] (8) to (99.58578,83.56239);
\node[draw=none,fill=none,scale=\labelscale] () at (102.64996,86.13354) {2};
\draw [black,->] (2) to (122.16004,44.46262);
\node[draw=none,fill=none,scale=\labelscale] () at (125.91881,45.83070) {2};
\draw [black,->] (7) to (130.00000,0.00000);
\node[draw=none,fill=none,scale=\labelscale] () at (134.00000,0.00000) {2};
\draw [black,->] (6) to (122.16004,-44.46262);
\node[draw=none,fill=none,scale=\labelscale] () at (125.91881,-45.83070) {2};
\draw [black,->] (1) to (99.58578,-83.56239);
\node[draw=none,fill=none,scale=\labelscale] () at (102.64996,-86.13354) {2};
\draw [black,->] (19) to (65.00000,-112.58330);
\node[draw=none,fill=none,scale=\labelscale] () at (67.00000,-116.04740) {2};
\draw [black,->] (3) to (22.57426,-128.02501);
\node[draw=none,fill=none,scale=\labelscale] () at (23.26886,-131.96424) {2};
\end{tikzpicture}
          }

	\caption{A graph where the default drawing is not usable: the smallest plane triangulation without a spanning 2-tree. A drawing with one vertex placed at infinity is better and shows the rotational
        symmetry with axis through the two vertices of large degree.}   
	\label{fig:no2tree}
\end{figure}

In this article, a tool is presented that can help to produce drawings of maps in oriented surfaces. It is just a tool: if the drawing produced helps to
visualize a map and/or to save time preparing drawings for articles, the aim is reached. The drawings are not guaranteed to optimize any criteria, but rely on
some heuristic approach. In some -- in my experience exceptional and astonishingly few -- cases the drawings are not usable. Examples are the double wheels or
the smallest plane triangulation without a spanning 2-tree shown in Figure~\ref{fig:no2tree}. Nevertheless such isolated examples should not be a reason to
modify the whole approach and maybe slow down all the other cases or even make other results worse.

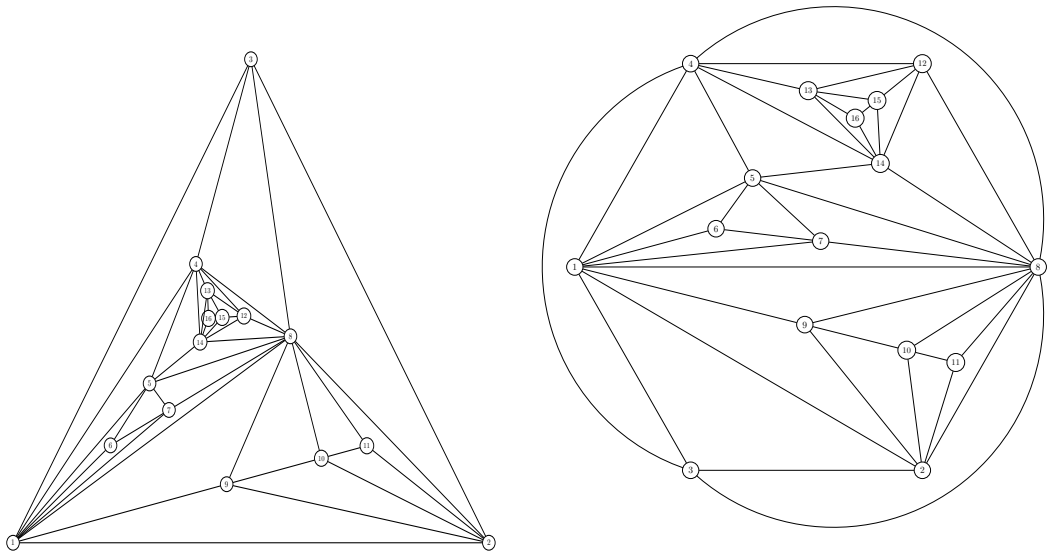
\begin{figure}[p]
	\centering
        \resizebox{0.45\textwidth}{0.45\textwidth}
        {
          \begin{tikzpicture}[scale=0.07]
\def\vertexscale{0.52}
\node [circle,black,draw,fill=white,scale=\vertexscale] (1) at (-86.60254,-50.00000) {1};
\node [circle,black,draw,fill=white,scale=\vertexscale] (2) at (86.60254,-50.00000) {2};
\node [circle,black,draw,fill=white,scale=\vertexscale] (3) at (-0.00000,100.00000) {3};
\node [circle,black,draw,fill=white,scale=\vertexscale] (4) at (-20.00842,36.49766) {4};
\node [circle,black,draw,fill=white,scale=\vertexscale] (5) at (-36.90439,-0.58062) {5};
\node [circle,black,draw,fill=white,scale=\vertexscale] (6) at (-51.11119,-19.78957) {6};
\node [circle,black,draw,fill=white,scale=\vertexscale] (7) at (-29.84072,-8.80224) {7};
\node [circle,black,draw,fill=white,scale=\vertexscale] (8) at (14.39052,14.04819) {8};
\node [circle,black,draw,fill=white,scale=\vertexscale] (9) at (-8.84026,-31.83630) {9};
\node [circle,black,draw,fill=white,scale=0.9*\vertexscale] (10) at (25.62530,-23.78626) {10};
\node [circle,black,draw,fill=white,scale=0.9*\vertexscale] (11) at (42.20697,-19.91383) {11};
\node [circle,black,draw,fill=white,scale=0.9*\vertexscale] (12) at (-2.54581,20.34896) {12};
\node [circle,black,draw,fill=white,scale=0.9*\vertexscale] (13) at (-15.81096,28.18820) {13};
\node [circle,black,draw,fill=white,scale=0.9*\vertexscale] (14) at (-18.49851,12.26547) {14};
\node [circle,black,draw,fill=white,scale=0.9*\vertexscale] (15) at (-10.46495,19.90264) {15};
\node [circle,black,draw,fill=white,scale=0.9*\vertexscale] (16) at (-15.45593,19.60209) {16};
\draw [black] (1) to (9);
\draw [black] (1) to (2);
\draw [black] (1) to (3);
\draw [black] (1) to (4);
\draw [black] (1) to (5);
\draw [black] (1) to (6);
\draw [black] (1) to (7);
\draw [black] (1) to (8);
\draw [black] (2) to (3);
\draw [black] (2) to (9);
\draw [black] (2) to (10);
\draw [black] (2) to (11);
\draw [black] (2) to (8);
\draw [black] (3) to (4);
\draw [black] (3) to (8);
\draw [black] (4) to (5);
\draw [black] (4) to (8);
\draw [black] (4) to (12);
\draw [black] (4) to (13);
\draw [black] (4) to (14);
\draw [black] (5) to (6);
\draw [black] (5) to (14);
\draw [black] (5) to (8);
\draw [black] (5) to (7);
\draw [black] (6) to (7);
\draw [black] (7) to (8);
\draw [black] (8) to (9);
\draw [black] (8) to (14);
\draw [black] (8) to (12);
\draw [black] (8) to (11);
\draw [black] (8) to (10);
\draw [black] (9) to (10);
\draw [black] (10) to (11);
\draw [black] (12) to (13);
\draw [black] (12) to (14);
\draw [black] (12) to (15);
\draw [black] (13) to (14);
\draw [black] (13) to (15);
\draw [black] (13) to (16);
\draw [black] (14) to (16);
\draw [black] (14) to (15);
\draw [black] (15) to (16);
\end{tikzpicture}
        }
              \resizebox{0.5\textwidth}{0.5\textwidth}
        {
          \begin{tikzpicture}[scale=0.07]
\def\vertexscale{0.80}
\tkzDefPoint(-50.00000,-86.60254){A}
\tkzDefPoint(-114.00000,-0.00000){B}
\tkzDefPoint(-50.00000,86.60254){C}
\tkzCircumCenter(A,B,C)\tkzGetPoint{D}
\tkzDrawArc[black](D,C)(A)
\tkzDefPoint(-50.00000,86.60254){A}
\tkzDefPoint(57.00000,98.72690){B}
\tkzDefPoint(100.00000,0.00000){C}
\tkzCircumCenter(A,B,C)\tkzGetPoint{D}
\tkzDrawArc[black](D,C)(A)
\tkzDefPoint(100.00000,0.00000){A}
\tkzDefPoint(57.00000,-98.72690){B}
\tkzDefPoint(-50.00000,-86.60254){C}
\tkzCircumCenter(A,B,C)\tkzGetPoint{D}
\tkzDrawArc[black](D,C)(A)
\node [circle,black,draw,fill=white,scale=\vertexscale] (1) at (-100.00000,-0.00000) {1};
\node [circle,black,draw,fill=white,scale=\vertexscale] (2) at (50.00000,-86.60254) {2};
\node [circle,black,draw,fill=white,scale=\vertexscale] (3) at (-50.00000,-86.60254) {3};
\node [circle,black,draw,fill=white,scale=\vertexscale] (4) at (-50.00000,86.60254) {4};
\node [circle,black,draw,fill=white,scale=\vertexscale] (5) at (-23.26571,37.87638) {5};
\node [circle,black,draw,fill=white,scale=\vertexscale] (6) at (-39.04894,16.29231) {6};
\node [circle,black,draw,fill=white,scale=\vertexscale] (7) at (6.11701,11.00022) {7};
\node [circle,black,draw,fill=white,scale=\vertexscale] (8) at (100.00000,0.00000) {8};
\node [circle,black,draw,fill=white,scale=\vertexscale] (9) at (-0.74004,-24.56047) {9};
\node [circle,black,draw,fill=white,scale=0.9*\vertexscale] (10) at (43.25308,-35.44594) {10};
\node [circle,black,draw,fill=white,scale=0.9*\vertexscale] (11) at (64.41769,-40.68283) {11};
\node [circle,black,draw,fill=white,scale=0.9*\vertexscale] (12) at (50.00000,86.60254) {12};
\node [circle,black,draw,fill=white,scale=0.9*\vertexscale] (13) at (0.73295,75.09452) {13};
\node [circle,black,draw,fill=white,scale=0.9*\vertexscale] (14) at (31.88349,44.08767) {14};
\node [circle,black,draw,fill=white,scale=0.9*\vertexscale] (15) at (30.42175,70.92158) {15};
\node [circle,black,draw,fill=white,scale=0.9*\vertexscale] (16) at (21.00150,63.38093) {16};
\draw [black] (1) to (9);
\draw [black] (1) to (2);
\draw [black] (1) to (3);
\draw [black] (1) to (4);
\draw [black] (1) to (5);
\draw [black] (1) to (6);
\draw [black] (1) to (7);
\draw [black] (1) to (8);
\draw [black] (2) to (3);
\draw [black] (2) to (9);
\draw [black] (2) to (10);
\draw [black] (2) to (11);
\draw [black] (2) to (8);
\draw [black] (4) to (12);
\draw [black] (4) to (13);
\draw [black] (4) to (14);
\draw [black] (4) to (5);
\draw [black] (5) to (6);
\draw [black] (5) to (14);
\draw [black] (5) to (8);
\draw [black] (5) to (7);
\draw [black] (6) to (7);
\draw [black] (7) to (8);
\draw [black] (8) to (11);
\draw [black] (8) to (10);
\draw [black] (8) to (9);
\draw [black] (8) to (14);
\draw [black] (8) to (12);
\draw [black] (9) to (10);
\draw [black] (10) to (11);
\draw [black] (12) to (13);
\draw [black] (12) to (14);
\draw [black] (12) to (15);
\draw [black] (13) to (14);
\draw [black] (13) to (15);
\draw [black] (13) to (16);
\draw [black] (14) to (16);
\draw [black] (14) to (15);
\draw [black] (15) to (16);
\end{tikzpicture}
          }

	\caption{When the outer face is a triangle, the surface of the inner face is smaller than for other polygons, which obviously forces vertices inside to be closer together. Option C allows the program to draw the outer edges in a curved way to create more room inside.}   
	\label{fig:straight_round}
\end{figure}

One -- but not the only -- problem for double wheels is that it is a triangulation. When the outer face is a triangle, the surface of the inner face is smaller
than for other polygons, which obviously forces vertices inside to be closer together. Option {\em C} allows the program to draw the outer edges in a curved way to
create more room inside like shown in Figure ~\ref{fig:straight_round}. Nevertheless  for the double wheels or the smallest plane triangulation without a spanning 2-tree
this does not help enough.

For these two cases an option implemented only for plane graphs and included to show rotational symmetry in case there is an axis going through two vertices can
be used: option {\em O}. It places one vertex at infinity.  The result is shown on the left of Figure~\ref{fig:no2tree}. For the double wheel and the graph in
Figure~\ref{fig:no2tree} one gets drawings that can be easily interpreted, but of course there are also other cases, where none of the options gives good
results.

\bibliographystyle{plain} \bibliography{../literatur.bib}

\begin{thebibliography}{10}

\bibitem{genuscomp}
G.~Brinkmann.
\newblock A practical algorithm for the computation of the genus.
\newblock {\em Ars Mathematica Contemporanea}, 22(4), 2022.
\newblock article \#P4.01.

\bibitem{allembed}
G.~Brinkmann.
\newblock Generating maps on oriented surfaces using the homomorphism
  principle.
\newblock arXiv article 2408.16512 \verb+https://arxiv.org/abs/2408.16512+, to
  appear in {\em Discrete \& Computational Geometry}, 2025.

\bibitem{BrDeDrHa97}
G.~Brinkmann, O.~Delgado Friedrichs, A.~Dress, and T.~Harmuth.
\newblock {C}a{G}e -- a virtual environment for studying some special classes
  of large molecules.
\newblock {\em {MATCH Commun. Math. Comput. Chem.}}, 36:233--237, 1997.
\newblock \verb+http://www.mathematik.uni-bielefeld.de/~CaGe+.

\bibitem{plantri}
G.~Brinkmann and B.D. McKay.
\newblock Fast generation of planar graphs.
\newblock {\em MATCH Commun. Math. Comput. Chem.}, 58(2):323--357, 2007.
\newblock see http://users.cecs.anu.edu.au/\~{}bdm/plantri/.

\bibitem{Olaf_graphzeichnen}
O.~{Delgado~Friedrichs}.
\newblock Fast embeddings for planar molecular graphs.
\newblock In P.~Hansen, P.~Fowler, and M.~Zheng, editors, {\em Discrete
  Mathematical Chemistry}, volume~51 of {\em DIMACS Series in Discrete
  Mathematics and Theoretical Computer Science}, pages 85--95. American
  Mathematical Society, 2000.

\bibitem{drawing_genus_2009}
C.A. Duncan, M.T. Goodrich, and S.G. Kobourov.
\newblock Planar drawings of higher-genus graphs.
\newblock In D.~Eppstein and E.R. Gansner, editors, {\em Graph Drawing 2009},
  volume 5849 of {\em Lecture Notes in Computer Science}, pages 45--56.
  Springer, 2010.

\bibitem{top_graph_theory}
J.L. Gross and T.W. Tucker.
\newblock {\em Topological Graph Theory}.
\newblock John Wiley and Sons, 1987.

\bibitem{torusdraw}
W.~Kocay, D.~Neilson, and R.~Szypowski.
\newblock Drawing graphs on the torus.
\newblock {\em Ars Combinatoria}, 59:259--277, 2001.

\bibitem{torusobstructions}
W.~Myrvold and J.~Woodcock.
\newblock A large set of torus obstructions and how they were discovered.
\newblock {\em Electronic Journal of Combinatorics}, 25(1), 2018.
\newblock P1.16.

\bibitem{sulanke}
T.~Sulanke.
\newblock Generating maps on surfaces.
\newblock {\em Discrete and Computational Geometry}, 57(2):335--356, 2017.

\bibitem{Tutte_draw}
W.T. Tutte.
\newblock How to draw a graph.
\newblock {\em Proc. London Math. Soc.}, 13:743--767, 1963.

\bibitem{arjana_draw}
A.~{\v Z}itnik.
\newblock Drawing graphs on surfaces.
\newblock {\em SIAM J.Disc.Math.}, 7(4):593--597, 1994.

\end{thebibliography}

\end{document}